\documentclass[a4paper,12pt]{article}
\usepackage[utf8]{inputenc}
\usepackage[T1]{fontenc}
\usepackage{epsfig,graphicx,xcolor,amsbsy,amssymb,latexsym,amsfonts,amsmath,tcolorbox,setspace,mathrsfs}
\usepackage{pstricks}
\usepackage{color}
\usepackage{soul}
\usepackage{accents}
\usepackage{dsfont}

\usepackage{simpler-wick}
\usepackage{placeins}
\usepackage{wrapfig,framed,caption}
\usepackage[makeroom]{cancel}
\usepackage{amsfonts}
\usepackage{dsfont}
\usepackage{mathrsfs}

\definecolor{shadecolor}{gray}{0.925}

\usepackage{eurosym}
\usepackage{enumitem}
\usepackage[a4paper]{geometry}
\usepackage{hyperref}
\usepackage{tcolorbox}
\usepackage{graphicx}
\usepackage{tikz}
\usetikzlibrary{cd}
\usetikzlibrary{decorations.pathreplacing,decorations.markings,snakes}
\usetikzlibrary{decorations.pathmorphing}
\usepackage{xcolor}
\usepackage{amsmath,amssymb,amsfonts,pstricks,setspace}
\usepackage[enableskew]{youngtab}
\usepackage{ytableau}

\numberwithin{equation}{section}

\newcommand{\bea}{\begin{eqnarray}\displaystyle}
\newcommand{\eea}{\end{eqnarray}}

\newcommand{\figref}[1]{Fig.~\protect\ref{#1}}

\newcommand{\quivbox}[3]{\draw[ thick,fill=white] (#1-#3,#2-#3) -- (#1-#3,#2+#3) -- (#1+#3,#2+#3) -- (#1+#3,#2-#3) -- (#1-#3,#2-#3);}

\newcommand{\tq}{\Tilde{q}_2}

\newcommand{\Qs}{Q_{S}}

\newcommand{\Qr}{Q_{\rho}}

\usepackage{float}
\newtcolorbox{summary}[2][]{colbacktitle=blue!10!white, colback=yellow!10!white,coltitle=blue!70!black, title={#2},fonttitle=\bfseries,#1}

\title{
{\bf Surface Defects in $A$-type Little String Theories}\\[40pt]}

\author{\large \textsc{Baptiste~Filoche\footnote{\tt b.filoche@ip2i.in2p3.fr}}~~,~~\textsc{Stefan~Hohenegger\footnote{\tt s.hohenegger@ipnl.in2p3.fr}}
~,~\,and\,~\textsc{Taro~Kimura\footnote{\tt taro.kimura@ube.fr}}
}

\begin{document}

\maketitle

\thispagestyle{empty}
\begin{center}
\renewcommand{\thefootnote}{\fnsymbol{footnote}}\vspace{-0.5cm}
${}^{\footnotemark[1]\,\footnotemark[2]}$ Universit\'e Claude Bernard Lyon 1, CNRS/IN2P3, IP2I Lyon, UMR 5822, Villeurbanne, F-69100, France\\[0.2cm]
${}^{\footnotemark[3]}$ Institut de Math\'ematiques de Bourgogne, Universit\'e Bourgogne Europe, CNRS, UMR 5584, France\\[2.5cm]
\end{center}

\begin{abstract}
$A$-type Little String Theories (LSTs) are engineered from parallel M5-branes on a circle $\mathbb{S}_\perp^1$, probing a transverse $\mathbb{R}^4/\mathbb{Z}_M$ background. Below the scale of the radius of $\mathbb{S}_\perp^1$, these theories resemble a circular quiver gauge theory with $M$ nodes of gauge group $U(N)$ and matter in the bifundamental representation (or adjoint in the case of $M=1$). In this paper, we study these LSTs in the presence of a surface defect, which is introduced through the action of a $\mathbb{Z}_N$ orbifold that breaks the gauge groups into $[U(1)]^N$. We provide a combinatoric expression for the non-perturbative BPS partition function for this system. This form allows us to argue that a number of non-perturbative symmetries, that have previously been established for the LSTs, are preserved in the presence of the defect. Furthermore, we discuss the Nekrasov-Shatashvili (NS) limit of the defect partition function: focusing in detail on the case $(M,N)=(1,2)$, we analyse two distinct proposals made in the literature. We unravel an algebraic structure that is responsible for the cancellation of singular terms in the NS limit, which we generalise to generic $(M,N)$. In view of the dualities of higher dimensional gauge theories to quantum many-body systems, we provide indications that our combinatoric expression for the defect partition are useful in constructing and analysing quantum integrable systems in the future.    
\end{abstract}

\newpage

\tableofcontents

\section{Introduction and Summary}

\subsection{Introduction}

A large number of interesting quantum theories can be constructed from String Theory, through limits that decouple the gravitational sector. Among the most peculiar classes of such theories are \emph{Little String Theories} (LSTs) \cite{Witten:1995zh,Aspinwall:1997ye,Intriligator:1997dh,Hanany:1997gh,Brunner:1997gf}: these theories behave like  quantum field theories (with ordinary point-like degrees of freedom) at low energies. However, depending on the details of their construction, they contain extended degrees of freedom at energies surpassing a specific scale. Due to their construction, these theories inherit many of the structures and symmetries of full String Theory (or its higher dimensional avatars), while at the same time being potentially free of the conceptual issues related to the gravitational sector. Such theories therefore constitute a unique arena to study many of the interesting features and conceptual aspects of String Theory in a simplified setting. This arena encompasses in particular many of the (geometric) techniques included in the toolbox of string model-building, such as compactifications or orbifolds. Moreover, the vast network of dualities of String Theory also leads to (surprising) relations among different LSTs or yet further classes of interesting theories, such as (quantum) integrable systems.

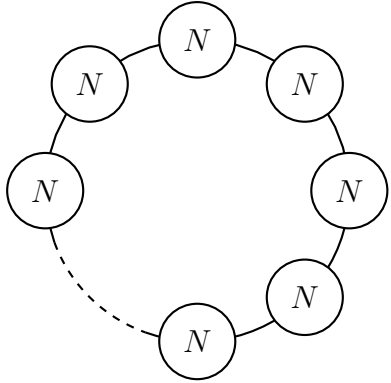
\begin{figure}[htbp]
    \centering
    \begin{tikzpicture}
        \draw [ thick,domain=0:200] plot ({2*cos(\x)}, {2*sin(\x)});
        \draw [ thick,dashed,domain=200:250] plot ({2*cos(\x)}, {2*sin(\x)});
        \draw [ thick,domain=250:360] plot ({2*cos(\x)}, {2*sin(\x)});
        \draw[ thick,fill=white] (2,0) circle (0.5cm);
        \node at (2,0) {$N$};
        \draw[ thick,fill=white] ({2*cos(45)}, {2*sin(45)}) circle (0.5cm);
        \node at ({2*cos(45)}, {2*sin(45)}) {$N$};
        \draw[ thick,fill=white] ({2*cos(90)}, {2*sin(90)}) circle (0.5cm);
        \node at ({2*cos(90)}, {2*sin(90)}) {$N$};
        \draw[thick,fill=white] ({2*cos(135)}, {2*sin(135)}) circle (0.5cm);
        \node at ({2*cos(135)}, {2*sin(135)}) {$N$};
        \draw[ thick,fill=white] ({2*cos(180)}, {2*sin(180)}) circle (0.5cm);
        \node at ({2*cos(180)}, {2*sin(180)}) {$N$};
        \draw[thick,fill=white] ({2*cos(270)}, {2*sin(270)}) circle (0.5cm);
        \node at ({2*cos(270)}, {2*sin(270)}) {$N$};
        \draw[thick,fill=white] ({2*cos(315)}, {2*sin(315)}) circle (0.5cm);
        \node at ({2*cos(315)}, {2*sin(315)}) {$N$};
    \end{tikzpicture}
    \caption{Cyclic quiver following the Dynkin diagram of $\widehat{A}_{M-1}$ and corresponding to the LST obtained by considering $N$ M$5$-branes on $\mathbb{R}^4 \times \mathbb{T}^2$ separated along $\mathbb{S}^1_{\perp}$ and probing a transverse $\mathbb{R}^4/\mathbb{Z}_M$ orbifold}
    \label{fig:Amquiver}
\end{figure}

In order to limit the playing field, in this work, we shall restrict ourselves to a class of LSTs with $\mathcal{N}=(2,0)$ supersymmetry (in six dimensions). Indeed, using methods similar as in the case of higher dimensional super-conformal field theories, such LSTs have been systematically classified \cite{Bhardwaj:2015oru,Bhardwaj:2019hhd,Bhardwaj:2022ekc} and here we shall deal specifically with the so-called $A$\emph{-type LSTs}. Due to the fact that these theories can be constructed using a number of complementary methods \cite{Haghighat:2013gba,Hohenegger:2013ala,Haghighat:2013tka}, these theories have been studied extensively in recent years~\cite{Hohenegger:2015btj,Hohenegger:2015cba,Hohenegger:2016yuv,Hohenegger:2016eqy,Ahmed:2017hfr,Haghighat:2017vch,Bastian:2017ing,Bastian:2017ary,Bastian:2017jje,Kimura:2017hez,Bastian:2018fba,Haghighat:2018dwe,Haghighat:2018gqf,Bastian:2018dfu,Bastian:2018jlf,Bastian:2019hpx,Bastian:2019wpx,Hohenegger:2019tii,Hohenegger:2020gio,Hayashi:2021pcj,Hohenegger:2020slq,Kimura:2022zsm,Wei:2022hjx,Filoche:2023yfm,Filoche:2024knd} and numerous highly non-trivial structures, symmetries and dualities have been uncovered \cite{Bastian:2017ary,Bastian:2018dfu,Ahmed:2017hfr,Bastian:2018jlf,Bastian:2019hpx,Bastian:2019wpx,Filoche:2023yfm}: a common construction of these theories is as the (six-dimensional) world-volume theory of $N$ parallel M5-branes separated along a circle $\mathbb{S}_\perp^1$ (see \cite{Haghighat:2013gba,Hohenegger:2013ala,Haghighat:2013tka,Hohenegger:2015cba}). This theory admits a number of dual descriptions \cite{Hohenegger:2013ala,Kanazawa:2016tnt,Hohenegger:2016yuv}. Here we shall focus on one, which at low energies (relative to the radius of $\mathbb{S}_\perp^1$) resembles a supersymmetric gauge theory on $\mathbb{R}^4 \times \mathbb{T}^2$ with a $U(N)$ gauge group and matter in the adjoint representation. We shall refer to this theory in the following as the $\widehat{A}_0$ theory. This theory has been generalised in \cite{Hohenegger:2016eqy} (see also \cite{Hohenegger:2013ala,Haghighat:2013tka}) to $\widehat{A}_{M-1}$ theories (also called Little String Orbifolds), by replacing the (flat) transverse space of the M5-branes with a $\mathbb{R}^4/\mathbb{Z}_M$ orbifold background. This generalises the low energy theory to a circular quiver gauge theory with $M$ gauge nodes of type $U(N)$ and hypermultiplet matter in the bifundamental representation (see Figure~\ref{fig:Amquiver}).

The full non-perturbative BPS partition function of the $\widehat{A}_{M-1}$ theory can be computed using a number of different techniques, exploiting various dual descriptions and computational techniques developed in String Theory. Moreover, exploiting notably a dual description of these theories as F-theory compactified on a class of toric Calabi-Yau threefolds \cite{Kanazawa:2016tnt}, a number of highly non-trivial symmetries have been established for the $\widehat{A}_M$, many of which act in an inherently non-perturbative fashion~\cite{Bastian:2018jlf,Filoche:2023yfm}. Furthermore, the double-elliptic fibration structure of the Calabi-Yau threefolds also implies two modular symmetries, with independent modular parameters. In particular for the case of the $\widehat{A}_0$ theory, these have been argued in \cite{Bastian:2019hpx,Bastian:2019wpx,Filoche:2023yfm} to be part of suitable paramodular groups. 

In this work we shall generalise the $A$-type LSTs and their orbifold extensions in yet a further way: indeed, we shall consider the theory in the presence of a codimension-2 surface defect. In the case of the $\widehat{A}_0$ theory, the latter is implemented by a $\mathbb{Z}_s$ orbifold (with $s\leq N$), that acts along a $\mathbb{C}$-plane in the world-volume of the M5-brane and a $\mathbb{C}$-plane along the transverse space. Its main effect is to break the $U(N)$ gauge group and we shall mostly be interested in the so-called \emph{full type} surface defect (with $s=N$), corresponding to $U(N)\to U(1)^N$. This defect can also be readily extended to the orbifolded case. There are several motivations for studying these theories in the presence of a defect:
\begin{itemize}
\item[\emph{(i)}] Such theoretical settings are interesting in their own right, providing access to a yet larger class of quantum theories. Moreover, due to the fact that they are constructed in a well-defined fashion from well studied theories, allows to extend many of the established computational techniques (after a slight adaptation), \emph{e.g.} ADHM constructions \cite{Kanno:2011fw,Lee:2020hfu,Chen:2019vvt}, algebraic methods \cite{Kimura:2022zsx,Kimura:2022zsm} or blow-up equations \cite{Kim:2023glm,Jeong:2020uxz}. In this way, we expect to access and study these theories in a fairly explicit and concrete fashion. Notably, as we shall demonstrate in this paper, it is possible to study non-perturbative aspects of the defect theories, by extending the analysis of symmetries of the $\widehat{A}_{M-1}$ theories mentioned above.
\item[\emph{(ii)}] Particular limits of the parameters of the $\widehat{A}_{M-1}$ theory with defect lead to (lower dimensional) supersymmetric gauge theories (see also \cite{Bastian:2018fba}). In particular, dimensionally reducing along either of the modular parameters mentioned above, successively replaces elliptic structures with trigonometric and rational ones. 
An overview of this situation is schematically provided in Table~\ref{tab:rtertesystems}, where columns and rows correspond to reductions of the two elliptic structures separately. For each combination, the red boxes specify the resulting supersymmetric gauge theory. A priori, these theories are intrinsically related to one another, and parts of the (non-perturbative) symmetries of the LSTs percolate to the lower dimensional theories, 
 which can thus be explored systematically by studying the defect LSTs. As we shall discuss in this article, among the theories that can be accessed in this manner we encounter one which closely resembles the Class $\mathcal{S}_k$ theories studied in \cite{Mitev:2017jqj,Bourton:2017pee,Bourton:2020rfo,Chen:2020jla}.

\item[\emph{(iii)}] The supersymmetric gauge theories obtained from the $A$-type LSTs can be related to various integrable models (see \cite{Donagi:1995cf,Braden:2001yc,Koroteev:2018isw,Bullimore:2014awa,Jeong:2021rll}). The precise type of theory is indicated by the blue boxes in Table~\ref{tab:rtertesystems}. Here, the Calogero-Moser systems have been discussed in \cite{Chen:2019vvt,Kimura:2022zsx,Grekov:2023fek,Kimura:2024kor}, the Ruijsenaars-Schneider systems in \cite{shiraishi2019affinescreeningoperatorsaffine,Kim:2024mnp} and the DELL system has been studied in \cite{Aminov:2014wra,Koroteev:2019gqi,Awata:2019isq,Gorsky:2021wio,Awata:2020yxf,Mironov:2023rzb}. Concretely, Higgsing the (non-perturbative) BPS partition function of the defect theory, is expected to lead to the wave function of the corresponding integrable system. Calculating an explicit expression of the latter is therefore essential in constructing the associated Hamilton operator, which is currently not known in many cases.
\end{itemize}

\begin{table}[htbp]
    \centering
    \begin{tabular}{c||c|c|c}
         & \hspace{1.41cm}\text{rational} \hspace{1.41cm} & \text{trigonometric} & elliptic  \\
         \hline \hline
         &&\\[-14pt]
          r. & \footnotesize{\hspace{-0.41cm}$\begin{matrix}\colorbox{red!50!white}{\parbox{4.5cm}{
              $\mathcal{N}=(2,2)$\text{ quiver th. on }$\mathbb{R}^2$}}\\ \vspace{-0.425cm} \\
              \colorbox{blue!25!white}{\parbox{4.5cm}{\text{Calogero-Moser system}}}
          \end{matrix}$\hspace{-0.4cm}} 
          & 
          \footnotesize{\hspace{-0.3cm}$\begin{matrix}\colorbox{red!50!white}{\parbox{4.75cm}{
              $\mathcal{N}=(2,2)^*$\text{ quiver th. on }$\mathbb{R}^2$}}\\ \vspace{-0.425cm} \\
               \colorbox{blue!25!white}{\parbox{4.75cm}{\text{t-Calogero-Moser system}}}
          \end{matrix}$\hspace{-0.35cm}} 
          &
           \footnotesize{\hspace{-0.15cm}$\begin{matrix}\colorbox{red!50!white}{\parbox{5cm}{
              $\mathcal{N}=2^*$\text{ quiver th. on }$\mathbb{R}^4$}}\\ \vspace{-0.425cm} \\
                \colorbox{blue!25!white}{\parbox{5cm}{\text{e-Calogero-Moser system}}}
          \end{matrix}$}\\
           \hline
           t.& \footnotesize{\hspace{-0.41cm}$\begin{matrix}\colorbox{red!50!white}{\parbox{4.5cm}{
            $\mathcal{N}=2$\text{ quiver th. on }$\mathbb{R}^2\times\mathbb{S}^1$}}\\ \vspace{-0.425cm} \\
              \colorbox{blue!25!white}{\parbox{4.5cm}{
            \text{Ruijsenaars-Schneider system}}}
           \end{matrix}$\hspace{-0.41cm}} 
           &
            \footnotesize{\hspace{-0.15cm}$\begin{matrix}\colorbox{red!50!white}{\parbox{4.75cm}{
            $\mathcal{N}=2^*$\text{ quiver th. on }$\mathbb{R}^2\times\mathbb{S}^1$}}\\ \vspace{-0.425cm} \\
             \colorbox{blue!25!white}{\parbox{4.75cm}{\text{t-Ruijsenaars-Schneider system}}}
           \end{matrix}$\hspace{-0.3cm}
           }
           &
           \footnotesize{\hspace{-0.15cm}$\begin{matrix}\colorbox{red!50!white}{\parbox{5cm}{
            $\mathcal{N}=1^*$\text{ quiver th. on }$\mathbb{R}^4\times\mathbb{S}^1$}}\\ \vspace{-0.425cm} \\
              \colorbox{blue!25!white}{\parbox{5cm}{\text{e-Ruijsenaars-Schneider system}}}
           \end{matrix}$\hspace{-0.45cm}}\\
           \hline
           e. & \footnotesize{\hspace{-0.41cm}$\begin{matrix}\colorbox{red!50!white}{\parbox{4.5cm}{
            $\mathcal{N}=1$\text{ quiver th. on }$\mathbb{R}^2\times\mathbb{T}^2$}}\\ \vspace{-0.425cm} \\
           \colorbox{blue!25!white}{\parbox{4.5cm}{ \text{dual e-CM system}}}
           \end{matrix}$\hspace{-0.41cm}} 
           &
            \footnotesize{\hspace{-0.3cm}$\begin{matrix}\colorbox{red!50!white}{\parbox{4.75cm}{
            $\mathcal{N}=1^*$\text{ quiver th. on }$\mathbb{R}^2\times\mathbb{T}^2$}}\\ \vspace{-0.425cm} \\
              \colorbox{blue!25!white}{\parbox{4.75cm}{\text{dual e-RS system}}}
           \end{matrix}$\hspace{-0.35cm}} 
           & \footnotesize{\hspace{-0.15cm}$\begin{matrix}\colorbox{red!50!white}{\parbox{5cm}{
            A\text{-type LST on }$\mathbb{R}^4\times\mathbb{T}^2$}}\\ \vspace{-0.425cm} \\
              \colorbox{blue!25!white}{\parbox{5cm}{\text{DELL system}}}
           \end{matrix}$}
    \end{tabular}
    \caption{Table summarising the different connections between supersymmetric gauge theories related to $A$-type LSTs and quantum integrable systems~\cite{Bullimore:2014awa}.}
    \label{tab:rtertesystems}
\end{table}

\noindent
Our main object of interest in this work is therefore the non-perturbative BPS \emph{defect partition function} of the $\widehat{A}_0$ and $\widehat{A}_{M-1}$ theories, denoted $\mathcal{Z}_{\rm def.}^{\rm inst.,(N)}$ and $\mathcal{Z}_{\rm def.}^{\rm inst.,(N,M)}$ respectively. While the former case has previously been discussed in the literature \cite{Koroteev:2019gqi,Awata:2020yxf}, to our knowledge, an explicit expression for the partition function in the case $M>1$ constitutes a genuinely new result. Moreover, in both cases, we provide a compact combinatoric expression for $\mathcal{Z}_{\rm def.}^{\rm inst.,(N,M)}$ as a function of all physical parameters of the defect theory: indeed, due to the orbifold action (that breaks the gauge group), the instanton counting parameter(s) of the LST is split into multiple parameters and we provide explicit expressions for the contribution of each of these instanton sectors in the partition function. We argue that the non-perturbative symmetries of the $\widehat{A}_0$ LSTs remain unbroken in the presence of the defect and conjecture a similar result also in the $\widehat{A}_{M-1}$-case.

Furthermore,  as in the case of the LST partition function $\mathcal{Z}_{\rm def.}^{\rm inst.,(N)}$ and $\mathcal{Z}_{\rm def.}^{\rm inst.,(N,M)}$ depend on two deformation parameters $\varepsilon_{1,2}$ that act as regulators in the sense of an $\Omega$-background \cite{Nekrasov:2003rj}. Indeed, $\mathcal{Z}_{\rm def.}^{\rm inst.,(N)}$ (as well as $\mathcal{Z}_{\rm def.}^{\rm inst.,(N,M)}$) is singular in the limit when one of these parameters vanishes (Nekrasov-Shatashvili (NS) limit \cite{Nekrasov:2009rc,Mironov:2009uv}). In the literature two different regularised partition functions have been conjectured (\cite{Lee:2020hfu,Koroteev:2019gqi,Kimura:2022zsx} and \cite{shiraishi2019affinescreeningoperatorsaffine,Mironov:2023rzb}), by normalising the partition function by a suitable quantity to remove the singular terms.  The regularity of theses proposals has been tested to limited order in the instanton expansion \cite{Koroteev:2019gqi,shiraishi2019affinescreeningoperatorsaffine} and both have a priori different physical properties. This last point is of particular interest in view of interpreting the regularisation of $\mathcal{Z}_{\rm def.}^{\rm inst.,(N)}$ as the eigenfunction of an integrable system. In this paper, we therefore analyse in detail the regularisation of the defect partition function of the $\widehat{A}_0$ theory, in the case of $N=2$: using the combinatoric form of the partition function, we unravel algebraic structures that guarantee the cancellation of poles in the regularised version. In this way we argue that both proposals in the literature lead to viable NS-limits and we highlight the difference among the two in terms of physical contributions at different instanton sectors. We indicate how these algebraic structures can be generalised to $N>2$ and we conjecture a relation to the blow-up equation \cite{Jeong:2020uxz}.

As first step towards analysing integrable structures, we consider the so-called bulk decoupling limit of the $N=2$ defect partition function of the $\widehat{A}_0$ theory: this corresponds to the vanishing of the diagonal coupling constant in the breaking of $U(2)\to U(1)\times U(1)$ and physically corresponds to reducing the instanton partition function to that of a vortex partition function living on the world-volume of the surface defect \cite{Koroteev:2019byp}. We find that the remaining expression can be rewritten in the form of an elliptic hypergeometric function, which is the eigenfunction of a specific (elliptic) operator, that is interpreted as the Hamiltonian.

\subsection{Summary of results}\label{sec:summary}
In order to render our results more readable, we provide a short summary in this Subsection. Indeed, the main results of this work are as follows:

\begin{enumerate}
\item {\bf Defect Partition Function for $\widehat{A}_0$ LSTs ($M=1$):}
We derive the defect partition function $\mathcal{Z}_{\rm def.}^{\rm inst.,(N)}$ for $\widehat{A}_0$ LSTs using an ADHM construction. The full non-perturbative contribution is expressed combinatorially as (see eq.~\eqref{eq:defZ}):
\begin{align}
    \mathcal{Z}_{\rm def.}^{\rm inst.,(N)} \left(\underline{\frak q},\mathfrak{q};\underline{Q_a},\Qr;\Qs,q_1,\tq\right) = \sum_{\underline{\lambda}\in \mathcal{P}^N} \prod_{i=1}^N \mathfrak{q}_i^{k_i(\underline{\lambda})} \prod_{1\leq i,j\leq N} \frac{\mathcal{N}_{\lambda^{(i)}\lambda^{(j)}}^{(j-i|N)}(\Qs Q_{a_j}/Q_{a_i},\Qr;q_{1},\tq)}{\mathcal{N}_{\lambda^{(i)}\lambda^{(j)}}^{(j-i|N)}(Q_{a_j}/Q_{a_i},\Qr;q_{1},\tq)},\label{ZdefN}
\end{align}
\noindent
where $(\underline{\frak q},\mathfrak{q};\underline{Q_a},\Qr;\Qs)$ label the physical moduli of the theory, while $q_1=e^{2\pi i\epsilon_1}$ and $\tilde{q}_2=e^{\frac{2\pi i\epsilon_2}{N}}$ denote the previously mentioned deformation parameters. Furthermore, $\mathcal{N}^{(p|N)}_{\mu\nu}$ are fractional Nekrasov subfunctions defined in (\ref{eq:Np}) and the summation $\underline{\lambda}=(\lambda^{(1)},\ldots,\lambda^{(N)})$ is over $N$ integer-partitions (see Section~\ref{sec:M=1} for more details on the notation). Our result is therefore fully combinatorical and
matches the partition function obtained in~\cite{shiraishi2019affinescreeningoperatorsaffine} from a vertex operator algebraic approach.
In the Nekrasov-Shatashvili (NS) limit $\epsilon_2\to 0$, the properly normalized defect partition function is conjectured to be the eigenfunction of the DELL system~\cite{Koroteev:2019gqi}.

\item  {\bf Defect Partition Function for $\widehat{A}_{M-1}$ LSTs ($M>1$):} We generalise the defect partition function for $\widehat{A}_{M-1}$ LSTs to include a $\mathbb{Z}_M$ orbifold. The corresponding ADHM quiver is given by a doubly periodic quiver which is shown in  Figure~\ref{fig:orrbichainsaw}. Similar to our results for the $\widehat{A}_0$ theory, the non-perturbative contribution is given by:
\begin{align}
    \mathcal{Z}_{\rm def.}^{\rm inst.,(N,M)}(\underline{\underline{\mathfrak{q}}}&,\mathfrak{q}|\underline{\underline{Q_a}},\Qr|Q_{\widehat{S}},q_1,\tq)=\sum_{\underline{\underline{\lambda}}\in \mathcal{P}^{NM}} \left[\prod_{i=1}^N \prod_{j=1}^M \left(\mathfrak{q}_{i,j} \right)^{k_i^j(\underline{\underline{\lambda}})} \right] \nonumber\\
    &\times\prod_{l=1}^M \prod_{1\leq i,j\leq N} \frac{\mathcal{N}_{\lambda^{(i,l-1)}\lambda^{(j,l)}}^{(j-i|N)}(Q_{\widehat{S}} Q_{a_j^l}/Q_{a_i^{l-1}},\Qr;q_{1},\tq)}{\mathcal{N}_{\lambda^{(i,l)}\lambda^{(j,l)}}^{(j-i|N)}(Q_{a_j^l}/Q_{a_i^{l}},\Qr;q_{1},\tq)}.\label{ZdefNM}
\end{align}
which uses a generalisation of the notation used in the $M=1$ case (we refer to Section~\ref{sec:genM} for the detailed notations).
To our knowledge, this result provides the first explicit expression for the defect partition function of $\widehat{A}_{M-1}$ LSTs.
In the NS limit, the normalized defect partition function is expected to be the eigenfunction of a spin generalisation of the DELL system~\cite{Koroteev:2019gqi}.

\item {\bf Nekrasov-Shatashvili (NS) Limit:} For general $N,M\in\mathbb{N}$, the defect partition function is singular in the NS limit ($\varepsilon_2 \to 0$). A finite limit can be obtained with the help of a suitable regularisation:
\begin{align}
\mathcal{Z}_{\rm def.}^{\rm inst.}\longrightarrow \frac{\mathcal{Z}_{\rm def.}^{\rm inst.}}{R}\,.
\end{align}
Following the literature, we consider two choices for the regulator $R$:
\begin{enumerate}
    \item Bulk normalization \cite{Lee:2020hfu,Koroteev:2019gqi,Kimura:2022zsx}: see \eqref{eq:ZbulkM}
    \item Shiraishi normalization and \cite{shiraishi2019affinescreeningoperatorsaffine,Mironov:2023rzb}): see \eqref{eq:ZfrakM}
\end{enumerate}
which correspond to different restrictions of the summation of integer partitions $\underline{\lambda}$ in (\ref{ZdefN}) and (\ref{ZdefNM}).
%
%
%
In particular, we elucidate the mechanism of pole cancellation in the combinatorial expression of the defect partition function, which is guaranteed by the following regularity condition:
\begin{align}
    \mathcal{Z}_{\rm def.}^{\rm inst.}[\underline{\lambda}] - \sum_{\underline{\nu} \subset \mathcal{S}(\underline{\lambda})} \mathcal{Z}_{\rm def.}^{\rm inst.}[\underline{\nu}] \cdot \mathcal{Z}_{\rm bulk}^{\rm inst.}[\underline{\lambda}\ominus \underline{\nu}] = \mathcal{O}(\varepsilon_2^0),
\end{align}
\noindent
where $\underline{\lambda}$ denotes a certain tuple of integer partitions, $\mathcal{S}(\underline{\lambda})$ denotes a subset of tuples 
which are obtained from $\underline{\lambda}$ by removing certain boxes, and $\ominus$ denotes an operation which removes boxes from a partition.
These notations are explicitly defined for general $N,M$ by~\eqref{eq:defSN} and~\eqref{eq:minusoperation}.

\item {\bf Bulk Decoupling Limit:} In the bulk decoupling limit ($\mathfrak{q} \to 0$), the defect partition function reduces to a vortex partition function on $\mathbb{C} \times \mathbb{T}^2$. In the NS limit, such functions typically become eigenfunctions of the dual elliptic Ruijsenaars-Schneider system~\cite{Bullimore:2014awa,Koroteev:2019gqi}.
We briefly discuss this fact in Section~\ref{Sect:BulkHamilton} and~\ref{sec:bulkdecouplingN}, but leave a more detailed analysis of this limit for future work.

\item {\bf Non-Perturbative Symmetries:} As discussed in Section~\ref{subsec:npsymNM1} and~\ref{subsec:npsymNM}, the defect partition function inherits non-perturbative symmetries from the parent LSTs~\cite{Bastian:2018jlf,Filoche:2023yfm}, including:
\begin{itemize}
    \item Cyclic permutations of Coulomb branch moduli and fractional couplings.
    \item Symmetry under $S \to -S + \varepsilon_1 + \varepsilon_2/N$ corresponding to flop transitions in the geometry.
    \item Quasi-periodicity under $S \to S - \rho$ and $S \to S-\tau$ mixing the bifundamental mass parameter $S$ with the complexified gauge coupling $\tau$ and the affinisation parameter~$\rho$.
\end{itemize}

These results provide a comprehensive framework for studying $A$-type LSTs with surface defects and their connections to integrable systems.
\end{enumerate}

\noindent
Concretely, the remainder of this paper is organised as follows: In Section~\ref{sec:M=1}, we discuss the construction of the defect partition function for the $\widehat{A}_0$ LST, based on the world-volume theory 
of parallel M5-branes probing a flat transverse background. We provide a combinatoric expression of the defect partition 
function which encapsulates the full non-perturbative sector and investigate non-perturbative symmetries inherited from the LST. 
Furthermore, we study the regularity of NS limits of this partition function as well as further limits that are interesting from the 
perspective of integrability. In Section~\ref{sec:genM}, we generalise the results of Section~\ref{sec:M=1} to $\widehat{A}_{M-1}$ LSTs 
in the presence of a defect by considering a double orbifold brane construction. Finally, Section~\ref{sec:conclusion} contains our 
conclusions and directions for further research. This work is supplemented by two Appendices: Appendix~\ref{sec:def} collects mathematical 
definitions used throughout the main text and Appendix~\ref{sec:voa} is dedicated to an alternative derivation of the defect partition function 
discussed in Section~\ref{sec:M=1} using a vertex operator algebraic approach.

\section{Defect partition function of $\widehat{A}_0$ LSTs}\label{sec:M=1}

\subsection{Partial orbifolding}

In this subsection we will detail the construction of BPS partition function (PF) in presence of a codimension-2 surface defect. We start with the aforementioned M-theoretic setup for $A$-type LSTs and consider $N$ M5-branes in the configuration given on the left of Table~\ref{tab:m5config}.
\begin{table}[htbp]
    \centering
    \begin{tabular}{c||c|c|c|c|c|c}
         & $\mathbb{C}_1$ &$\mathbb{C}_2$ & $\mathbb{T}^2$ & $\mathbb{S}^1_{\perp}$ & $\mathbb{C}_3$ & $\mathbb{C}_4$  \\
         \hline\hline
       $N$ M$5$  & -\,- & -\,- & -\,- & $\times$ & &\\
       $k$ M$2$ &  & & -\,- & - & & \\
     \end{tabular}
     \quad
     \begin{tabular}{c||c|c|c|c|c|c}
         & $\mathbb{C}_1$ &$\mathbb{C}_2/\mathbb{Z}_s$ & $\mathbb{T}^2$ & $\mathbb{S}^1_{\perp}$ & $\mathbb{C}_3$ & $\mathbb{C}_4/\mathbb{Z}_s$  \\
         \hline\hline
       $N$ M$5$  & -\,- & -\,- & -\,- & $\times$ & &\\
       $k$ M$2$ &  & & -\,- & - & & \\
     \end{tabular}
    \caption{On the left M$5$/M$2$-brane setup engineering a $\widehat{A}_0$ quiver rank $N$ LST, M$5$-branes are separated along the $\mathbb{S}^1_{\perp}$. On the right, the modified situation engineering a codimension-2 surface defect.}
    \label{tab:m5config}
\end{table}
In this case, the low energy quiver gauge theory correspond to a $\widehat{A}_0$ quiver with one $U(N)$ gauge node and matter in the adjoint representation~\cite{Haghighat:2013gba,Hohenegger:2015btj}.
We are interested in the non-perturbative sector of the defect PF. In order to regularise the IR-divergences associated with instanton contribution, we introduce a general $\Omega$-background~\cite{Nekrasov:2003rj} on all non-compact dimensions. The $\Omega$-background is explicitly realised by gauging a $U(1)_{\varepsilon_i}$ action parametrised by $\{\varepsilon_i\}_{i=1,2,3,4}$ on all $\mathbb{C}$ planes. In the following, we denote by $\mathbb C_i = \mathbb{C}_{\varepsilon_i}$ the complex plane with a $U(1)_{\varepsilon_i}$ isometry. In order to preserve at least $\mathcal{N}=(1,0)$ supersymmetry in six dimensions, one needs to impose that the isometry group is a subgroup of the group rotating the $\mathbb{C}^4$ planes: $U(1)_{\varepsilon_1} \times U(1)_{\varepsilon_2} \times U(1)_{\varepsilon_3} \times U(1)_{\varepsilon_4} \subset Spin(8)$~\cite{Nekrasov:2016gud}. This adds an extra constraint on the $\Omega$-background parameters $\{\varepsilon_i\}_{i=1,2,3,4}$:
\begin{align}
    \varepsilon_1+\varepsilon_2+\varepsilon_3+\varepsilon_4=0\,.
\end{align}
In the following, we make this constraint manifest and use the following parametrisation:
\begin{align}\label{eq:parameps}
    \varepsilon_3 = -S\,, && \varepsilon_4=S-\varepsilon_1-\varepsilon_2\,,
\end{align}
where $S$ is interpreted from the LST perspective, i.e. in the M$5$-branes world-volume theory, as the mass parameter of the adjoint matter.

\paragraph{}
The presence of a codimension-2 surface defect can be geometrically emulated by introducing a $\mathbb{Z}_s$ orbifold on the $\mathbb{C}_2$ and $\mathbb{C}_4$ planes \cite{Chen:2019vvt}. In coordinates, the orbifold action is given by
\begin{align}\label{eq:orbifoldactionM1}
    \mathbb{Z}_s : \begin{pmatrix}
        z_2\\z_4
    \end{pmatrix} \longrightarrow \begin{pmatrix}
        e^{2i \pi \frac{1}{s} } & 0 \\ 0 & e^{-2i\pi \frac{1}{s}}
    \end{pmatrix}\cdot \begin{pmatrix}
        z_2\\ z_4
    \end{pmatrix}&& \text{with} &&(z_1,z_2,z_3,z_4)\in \mathbb{C}_1 \times \mathbb{C}_2 \times \mathbb{C}_3 \times \mathbb{C}_4\,.
\end{align}
The orbifolded situation is given by the tabular on the right of Table~\ref{tab:m5config}. From the worldvolume of the M5-branes perspective, the orbifold breaks the gauge group according to
\begin{align}\label{eq:breakingpattern}
    U(N) \longrightarrow U(n_1) \times \cdots \times U(n_s)\,,&& \text{with} && n_1+\cdots+n_s = N\,,
\end{align}
where orbifold sub-sectors are classified by irreducible representations of $\mathbb Z_s$. In the particular case $s=N$, the gauge group is maximally broken to $U(1)^N$, which corresponds to the so-called {\it full type} surface defect \cite{Kanno:2011fw}. In the following we restrict ourselves to $s=N$, but the discussion can be naturally extended to $s<N$.

\subsection{Character computation}\label{subsec:characcomM1}
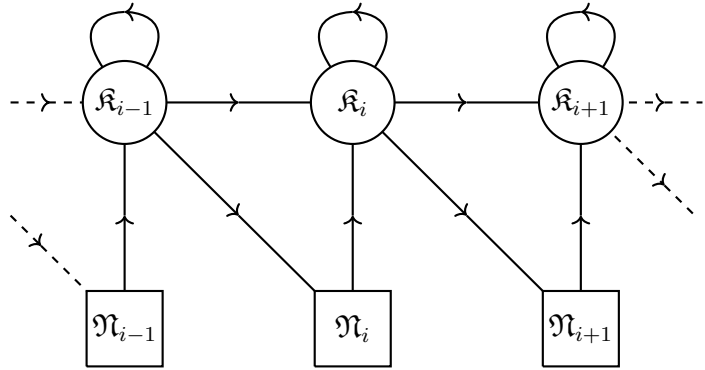
\begin{figure}[htbp]
    \centering
    \begin{tikzpicture}
        \draw[thick,->] (0,0.1) .. controls (1,1.2) and (0,1.2) .. (0,1.2);
        \draw[thick] (0,1.2) .. controls (0,1.2) and (-1,1.2) .. (0,0.1 );
        \draw[thick,->] (3,0.1) .. controls (4,1.2) and (3,1.2) .. (3,1.2);
        \draw[thick] (3,1.2) .. controls (3,1.2) and (2,1.2) .. (3,0.1 );
        \draw[thick,->] (-3,0.1) .. controls (-2,1.2) and (-3,1.2) .. (-3,1.2);
        \draw[thick] (-3,1.2) .. controls (-3,1.2) and (-4,1.2) .. (-3,0.1 );
        \draw[thick,->] (0,0) -- (1.5,0);
        \draw[thick,->] (1.5,0) -- (3,0);
        \draw[thick,->] (-3,0) -- (-1.5,0);
        \draw[thick,->] (-1.5,0) -- (0,0);
        \draw[thick,<-] (0,0) -- (0,-1.5);
        \draw[thick,<-] (0,-1.5) -- (0,-3);
        \draw[thick,<-] (3,0) -- (3,-1.5);
        \draw[thick,<-] (3,-1.5) -- (3,-3);
        \draw[thick,<-] (-3,0) -- (-3,-1.5);
        \draw[thick,<-] (-3,-1.5) -- (-3,-3);
        \draw[thick,->] (0,0) -- (1.5,-1.5);
        \draw[thick,->] (1.5,-1.5) -- (3,-3);
        \draw[thick,->] (-3,0) -- (-1.5,-1.5);
        \draw[thick,->] (-1.5,-1.5) -- (0,-3);
        \draw[dashed,thick] (-4.5,0) -- (-4.1,0);
        \draw[dashed,thick,>-] (-4.1,0) -- (-3,0);
        \draw[dashed,thick,->] (3,0) -- (4.1,0);
        \draw[dashed,thick] (4.1,0) -- (4.6,0);
        \draw[dashed,thick,->] (-4.5,-1.5) -- (-4.1,-1.9);
        \draw[dashed,thick,->] (-4.1,-1.9) -- (-3,-3);
        \draw[dashed,thick,->] (3,0) -- (4.1,-1.1);
        \draw[dashed,thick] (4.1,-1.1) -- (4.5,-1.5);
        \draw[ thick,fill=white] (0,0) circle (0.55cm);
        \draw[ thick,fill=white] (3,0) circle (0.55cm);
        \draw[ thick,fill=white] (-3,0) circle (0.55cm);
        \quivbox{0}{-3}{0.5}
        \quivbox{3}{-3}{0.5}
        \quivbox{-3}{-3}{0.5}
        \node at (0,0) {$\mathfrak{K}_i$};
        \node at (-3,0) {$\mathfrak{K}_{i-1}$};
        \node at (3,0) {$\mathfrak{ K}_{i+1}$};
        \node at (0,-3) {$\mathfrak{N}_i$};
        \node at (-3,-3) {$\mathfrak{N}_{i-1}$};
        \node at (3,-3) {$\mathfrak{N}_{i+1}$};
    \end{tikzpicture}
    \caption{A portion of the cyclic chain-saw ADHM quiver characterising the $\mathbb{Z}_N$-orbifolded instanton moduli space.}
    \label{fig:chainsaw}
\end{figure}

\paragraph{}
We now come to the construction of the defect BPS/instanton\footnote{We will use the term instanton to designate what corresponds in our case to BPS states since our setup admits dual descriptions in which M$2$-branes ending on M$5$-branes configurations are dual to instantons.} PF. We describe the construction in two parts: first, we review the ADHM construction of the character of the tangent space to the instanton moduli space on the background with a partial orbifold \cite{Kanno:2011fw,Lee:2020hfu,Chen:2019vvt,Kimura:2022zsx,Kimura:2022zsm}. This quantity can then be directly related to the defect instanton PF, as detailed thereafter. In the next subsection, we propose a combinatorial re-writing of the defect instanton PF which allows to easily relate our result to the $A$-type LST instanton PF. 
\paragraph{}
We start with several definitions which allow to characterise the instanton moduli space. We first define the framing vector space as:
\begin{align}
    \mathfrak{N} = \bigoplus_{i=1}^N \mathfrak{N}_i \otimes \mathfrak{R}_i\,, && {\rm dim}_{\mathbb{C}} \,\mathfrak{N}_i =N\,,
\end{align}
capturing the information of the M$5$-brane moduli. $\{\mathfrak{R}_i\}_{i\in \{1\ldots N\}}$ are irreducible representations of the orbifold group $\mathbb{Z}_N$. The splitting of the framing vector space follows the breaking pattern of the gauge group~\eqref{eq:breakingpattern}. For an instanton configuration of charge $k$ of the original $U(N)$ group, the instanton vector space (which characterises instanton moduli) splits into $N$ sectors and is given by:
\begin{align}
    \mathfrak{K} = \bigoplus_{i=1}^N \mathfrak{K}_i \otimes \mathfrak{R}_i\,,&& {\rm dim}_{\mathbb{C}}\,\mathfrak{K}_i = k_i\,.
\end{align}
We have $k=k_1 +   \ldots + k_N$ and we interpret the set of instanton charges $\{k_i\}_{i\in \{1,\ldots,N\}}$ as the instanton charges of each sub-$U(1)$ gauge group obtained from the breaking~\eqref{eq:breakingpattern}. Finally we define the co-tangent space to the space-time at the origin, \emph{i.e.} at the fixed point of the $U(1)_{\varepsilon_i}$ isometries, as:
\begin{align}
    &\mathfrak{Q} = T_o^\vee (\mathbb{C}_1 \times \mathbb{C}_2/\mathbb{Z}_N \times \mathbb{C}_3 \times \mathbb{C}_4/\mathbb{Z}_N)= \mathfrak{Q}_1 \oplus \mathfrak{Q}_2 \oplus \mathfrak{Q}_3 \oplus \mathfrak{Q}_4 \,.
\end{align}
By definition, the characters of the $\mathfrak{N}$ and $\mathfrak{K}$ vector spaces can be related respectively to M$5$-brane moduli $\{a_{i,\alpha}\}_{i,\alpha \in \{1,\ldots,N\}}$ and M$2$-brane moduli $\{\phi_{i,I}\}_{i\in\{1,\ldots,N\},I\in\{1,\ldots,k_i\}}$:
\begin{align}
    &{\bf N}_i:={\rm ch}\, \mathfrak{N}_i = \sum_{\alpha=1}^Ne^{2i\pi a_{i,\alpha}}\,, && i \in \{1, \ldots,N\}\,, \\
    &{\bf K}_i:={\rm ch}\, \mathfrak{K}_i = \sum_{I=1}^{k_i} e^{2i \pi \phi_{i,I}}\,, && i \in \{1, \ldots, N \}\,,
\end{align}
in general we use the notation: $Q_x:=e^{2i\pi x}$. Similarly, matrix-valued characters\footnote{We consider matrix-valued characters in order to keep the derivation compact.} of $\mathfrak{Q}$ are related to the $\Omega$-background parameters $\{\varepsilon_i\}_{i=1,2,3,4}$ through:
\begin{align}
   &{\bf Q}_1:={\rm ch}\, \mathfrak{Q}_1 = q_1 \mathcal{R}_0\,, && {\bf Q}_2:={\rm ch}\, \mathfrak{Q}_2 = \tq \mathcal{R}_1\,,\\
    &{\bf Q}_3:={\rm ch}\, \mathfrak{ Q}_3 = q_3 \mathcal{R}_0\,, &&{\bf Q}_4:={\rm ch}\, \mathfrak{ Q}_4 = q_4 \mathcal{R}_{N-1}\,,
\end{align}
with,
\begin{align}
    q_1 := e^{2i\pi \varepsilon_1}\,, && \tq = e^{\frac{2i\pi \varepsilon_2}{N}}\,, && q_3=e^{2i\pi \varepsilon_3}\,, && q_4 = e^{2i\pi \varepsilon_4}= q_1^{-1} \tq^{-1} q_3^{-1}\,,
\end{align}
and $\{\mathcal{R}_i\}_{i\in \{1\ldots N\}}$ are matrix representations of $\{\mathfrak{R}_i\}_{i\in \{1\ldots N\}}$ given by:
\begin{align}
    [\mathcal{R}_1]_{kl} = \delta_{k+1,l}\,, && \forall p \in \mathbb{Z}\,,&& \mathcal{R}_p = \mathcal{R}_1^p\,, && \mathcal{R}_N = \mathcal{R}_0=\mathds{1}_N\,,
\end{align}
with cyclic indices of the Kronecker delta which are commuting matrices. In addition, we define some useful notations to easily manipulate the matrix-valued characters:
\begin{align}
    {\bf P}_i := \mathds{1} - {\bf Q}_i\,, && {\bf P}_{ij}:= {\bf P}_i {\bf P}_j\,, && \forall i,j \in \{1,\ldots,4\}\,,
\end{align}
we also define the dual character ${\bf X}^\vee$ of a matrix-valued character ${\bf X}$:
\begin{align}\label{eq:dualoperation}
    \text{for} \quad\, {\bf X} = \sum_{i} n_i e^{2i\pi x_i} \mathcal{R}_i\,, && \, {\bf X}^\vee = \sum_i n_i e^{-2i\pi x_i} \mathcal{R}_i^\vee\,\,,
\end{align}
with $\mathcal{R}_i^\vee = \mathcal{R}_{N-i}$. We also define the $5$d/$6$d index of a character as:
\begin{align}\label{eq:6dindex}
    \text{for} \quad \, {\bf W} = \sum_{i} n_i e^{2i\pi w_i} \,, && \mathbb{I}[{\bf W}]= \begin{cases}
        \prod_i \vartheta(Q_{w_i};\Qr)^{n_i}\quad \text{in 6d}\,,\\
        \prod_i(1-Q_{w_i})^{n_i}\quad \text{in 5d}\,.
    \end{cases}
\end{align}
where $\vartheta$ is a theta function defined in \eqref{eq:appthetedef} and $\rho$ is the affine extension parameter of the $U(N)$ node. The defect instanton PF can be computed as a path integral over the instanton moduli of the $6$d index of the character of the tangent space to the instanton moduli space $\chi_{\rm inst.}$. We construct $\chi_{\rm inst.}$ using the eleven dimensional background, for a detailed review of this procedure see \cite{Kimura:2022zsm}. We first define the character of the observable sheaf \cite{Kimura:2022zsm} given by the following contribution (the M$5$-branes are located at $z_3,z_4=0$, while the instantons are localized by the $U(1)_{\varepsilon_i}$ at the origin of all four coordinates):
\begin{align}
    {\bf Y}:= {\bf P}_{34} [{\bf N} - {\bf P}_{12} {\bf K}]\,.
\end{align}
We can then define the character of analogue of a vector multiplet contribution on the $11$d background:
\begin{align}\label{eq:generalisedvector}
    {\bf V}:= {\bf Y}^\vee ({\bf P}_1{\bf P}_2 {\bf P}_3 {\bf P}_4)^{-1} {\bf Y}\,.
\end{align}
Once we use that ${\bf Q}_1{\bf Q}_2{\bf Q}_3{\bf Q}_4 = \mathds{1}_N$, ${\bf V}$ splits into a character and its dual:
\begin{align}
    {\bf V} = {\bf v} + {\bf v}^\vee\,, && {\bf v}:= {\bf P}_3^\vee \left[ {\bf N}^\vee {\bf P}_{12}^{-1} {\bf N} - {\bf N}^\vee {\bf K} - {\bf Q}_1^\vee {\bf Q}_2^\vee {\bf K}^\vee {\bf N}+ {\bf P}_{12}^\vee {\bf K}^\vee {\bf K} \right]\,.
\end{align}
The character of interest is then given by the $\mathbb{Z}_N$ invariant sub-sector of ${\bf v}$:
\begin{align}\label{eq:chidef}
   &\chi := \chi_{\rm pert.} + \chi_{\rm inst.}\,, && \chi_{\rm pert.}:= \left[{\bf P}_3^\vee {\bf N}^\vee {\bf P}_{12}^{-1} {\bf N}\right]_{\mathbb Z_N}\,,\\
    &\chi_{\rm inst.}:= \left[{\bf P}_3(- {\bf N}^\vee {\bf K} - {\bf Q}_1^\vee {\bf Q}_2^\vee {\bf K}^\vee {\bf N}+ {\bf P}_{12}^\vee {\bf K}^\vee {\bf K} )\right]_{\mathbb{Z}_N}\,,
\end{align}
where we define the invariant sub-sector as the trace of the matrix-valued characters:
\begin{align}
    \left[ \sum_{i=1}^N c_i \mathcal{R}_i \right]_{\mathbb{Z}_N} = \frac{1}{N}{\rm Tr} \, \left[\sum_{i=1}^N c_i \mathcal{R}_i \right]\,, && \forall c_i \in \mathbb{C}\,.
\end{align}
We obtain:
\begin{align}
    \chi_{\rm inst.} = (1-q_3^{-1}) \sum_{i=1}^N \left[- {\bf N}_i^\vee {\bf K}_i - q_1^{-1} \tq^{-1} {\bf K}_{i-1}^\vee {\bf N}_i - (1-q_1^{-1})\tq^{-1} {\bf K}_{i-1}^\vee {\bf K}_i + (1-q_1^{-1}) {\bf K}_i^\vee {\bf K}_i \right]\,,
\end{align}
which matches the chain-saw quiver characterization of the instanton moduli space~\cite{Kanno:2011fw} through the ADHM equations of the quiver~\figref{fig:chainsaw}. 

\subsection{Combinatoric expression}~\label{subsec:combexprM=1}
We now detail the reasoning leading to the final combinatoric expression of the instanton PF in presence of a surface defect. The contributions to the PF of instanton configurations at order $\underline{k}=\left( k_1, \ldots, k_N \right)$ denoted $\mathcal{Z}^{\underline{k},N}_{\rm def.}$ are then given by the $6$d index~\eqref{eq:6dindex} of $\chi_{\rm inst.}$, which corresponds to the following Losev-Moore-Nekrasov-Shatashvili (LMNS) formula:
\begin{align}\label{eq:LMNS}
    \mathcal{Z}^{\underline{k},N}_{\rm def.} =\frac{1}{\underline{k}!} \frac{[S-\varepsilon_1]^{|\underline{k}|}}{[S]^{|\underline{k}|}[-\varepsilon_1]^{|\underline{k}|}}& \oint \prod_{i=1}^N \prod_{I=1}^{k_i} \mathrm{d}\phi_{i,I} \prod_{i,\alpha=1}^N \prod_{I=1}^{k_i} \frac{[S+\phi_{i,I}-a_{i,\alpha}]}{[\phi_{i,I}-a_{i,\alpha}]} \prod_{I=1}^{k_{i-1}} \frac{[S+a_{i,\alpha} - \phi_{i-1,I} -\varepsilon_{1}-\frac{\varepsilon_2}{N}]}{[a_{i,\alpha} - \phi_{i-1,I} -\varepsilon_{1}-\frac{\varepsilon_2}{N}]}\nonumber\\
    &\times \prod_{1\leq I\neq J\leq k_i} \frac{[\phi_{i,J}-\phi_{i,I}][S+\phi_{i,J}-\phi_{i,I}-\varepsilon_1]}{[\phi_{i,J}-\phi_{i,I}-\varepsilon_1][S+\phi_{i,J}-\phi_{i,I}]}\nonumber \\
    &\times \prod_{I=1}^{k_{i-1}} \prod_{J=1}^{k_{i}} \frac{[\phi_{i,J}-\phi_{i-1,I}-\varepsilon_1-\frac{\varepsilon_2}{N}][S+\phi_{i,J}-\phi_{i-1,I}-\varepsilon_{1}-\frac{\varepsilon_2}{N}]}{[\phi_{i,J}-\phi_{i-1,I}-\frac{\varepsilon_2}{N}][S+\phi_{i,J}-\phi_{i-1,I}-\frac{\varepsilon_2}{N}]}\,.
\end{align}
Here indices are understood with the cyclicity property $\phi_{i+N,I}=\phi_{i,I}$ and we used the following shorthand notations:
\begin{align}
    [x]:= \vartheta(Q_x;\Qr)\,, && \frac{1}{\underline{k}!} = \prod_{i=1}^{N} \frac{1}{k_i !}\,, && |\underline{k}|=\sum_{i=1}^N k_i\,.
\end{align}
The combinatoric expression of the PF is then given by the residues of the integral~\eqref{eq:LMNS}, which correspond to the fixed points of the instanton moduli space under the action of ${\rm GL}({\bf K}) \times {\rm GL}({\bf N}) \times {\rm GL}({\bf Q})$. In order to understand the pole structure of the integral~\eqref{eq:LMNS}, we need to specify an integration ordering. To preserve a symmetric formulation in all sub-sector of the orbifold we choose:
\begin{align}
    \oint \prod_{i=1}^N \prod_{I=1}^{k_i} {\rm d}\phi_{i,I} \longrightarrow \prod_{i=1}^N  \oint {\rm d} \phi_{i,\max(\{k_j\}_{j=1,...,N})} \ldots \prod_{i=1}^N \oint {\rm d} \phi_{i,1}\,.
\end{align}
With this choice of ordering, the pole structure of the integrand is governed by the relations~\cite{Kanno:2011fw}:
\begin{align}
    \phi_{i,1}=a_{i,\alpha}\,, && \phi_{i,I} = \phi_{i,J} + \varepsilon_1\quad \text{for} \quad  I>J\geq1\,, && \phi_{i,I}=\phi_{i+1,J} + \frac{\varepsilon_2}{N}\quad \text{for} \quad  I,J\geq1\,.
\end{align}
The result of the integration can be written as a summation over several configurations following the Jeffrey-Kirwan residue prescription~\cite{Benini:2013xpa}. Non-equivalent configurations are then classified by a $N$-tuple of partitions\footnote{Strictly speaking a given configuration is characterised by a $N^2$-tuple of partition (corresponding to the $N^2$ starting points given by $\phi_{i,1}=a_{i,\alpha}$ with $\alpha,i=1,\ldots,N$) with interlacing condition~\cite{Feigin2008YangiansAC}. Such configuration is in bijection with a $N$-tuple of partitions.} $\underline{\lambda}=(\lambda^{(1)}, \ldots ,\lambda^{(N)})$ such that:
\begin{align}\label{eq:chKi}
     {\bf K}_i \big|_{\underline{\lambda}}= \sum_{j=1}^N  Q_{a_{i-j+1}} q_2^{\lfloor \frac{i-j}{N} \rfloor} \tq^{i} \sum_{(m,nN+j) \in \lambda^{(i-j+1)}} q_1^{m-1} q_2^{n}\,,
\end{align}
where we use cyclic conventions on the indices of $a_{i+N}=a_i$ and $\lambda^{(i+N)}=\lambda^{(i)}$. The $N$-tuple of partitions $\underline{\lambda}$ is related to the charges of the instanton configuration $\underline{k}=(k_1,\ldots,k_N)$ by a colouring of $\underline{\lambda}$. We define the colouring $c_N(\Box)$ of a box belonging to the Young diagram of a given partition in the following manner:
\begin{align}
    \forall \Box=(m,n) \in \lambda^{(\alpha)}\,, \quad c_N(\Box) = (\alpha+m-1 \mod N) +1\,.
\end{align}
For instance, at $N=3$ the $3$-tuple of partitions $(\lambda^{(1)}=(3,2), \lambda^{(2)}=(2,2,1), \lambda^{(3)}=(5))$ has the following colouring~\cite{Kanno:2011fw}:
\begin{align}\label{eq:configN3}
    \lambda^{(1)}:\,
    \begin{ytableau}
        1&1&1\\
        2&2
    \end{ytableau}\,,
    &&
    \lambda^{(2)}:\,
    \begin{ytableau}
        2&2\\
        3&3\\
        1
    \end{ytableau}\,,
    &&
    \lambda^{(3)}:\,
    \begin{ytableau}
        3&3&3&3&3
    \end{ytableau}\,.
\end{align}
The relation between the instanton charges of each sub-sector $\underline{k}$ and the $N$-tuple of partitions is then given by:
\begin{align}\label{eq:colorf}
    k_i(\underline{\lambda}) = \mathrm{Card} \left( \left\{ \Box \in \lambda^{(\alpha)} \big| c_N(\Box)=i\,,\quad \alpha = 1,\ldots,N\right\} \right)\,,\quad \forall i \in \{1,\ldots,N\}\,,
\end{align}
\emph{i.e.} the number of boxes in the configuration $\underline{\lambda}$ with colouring $i$. For instance the configuration described by~\eqref{eq:configN3} corresponds to $k_1 =4$, $k_2=7$ and $k_3=4$. 
\paragraph{}
Previous combinatoric results have been found in~\cite{Koroteev:2019byp,Kimura:2022spi}. We now give a new presentation which allows to see the defect instanton PF as an extension of the PF without defect. Considering~\eqref{eq:chKi}, we can organise the character computation in power of $\tq$. We define a projection $\omega_{k|N}$ which selects in a Laurent series in $\tq$ terms that come with a power congruent to $k\!\mod N$:
\begin{align}\label{eq:defomega}
    \omega_{k|N} : \sum_{n\in\mathbb{Z}} c_n\tq^n \longrightarrow \sum_{n\in \mathbb{Z}} \Tilde{c}_{n} \tq^{n} \quad \text{with} \quad \Tilde{c}_n = \begin{cases}
        \Tilde{c}_n = c_n \quad \text{if} \quad n \equiv k \mod N,\\
        0 \quad \text{otherwise}\,.
    \end{cases}
\end{align}
This projection has many useful properties given in Appendix~\ref{subsec:propome}. Using $\omega_{k|N}$, we can simply re-write the character of the instanton vector space at the fixed point $\underline{\lambda}$:
\begin{align}
     {\bf K}_{i}\big|_{\underline{\lambda}} = \sum_{j=1}^N  {\bf K}_{i,j}\big|_{\underline{\lambda}}\,,
\end{align}
such that:
\begin{align}
     {\bf K}_{i,j}\big|_{\underline{\lambda}} := \begin{cases}
    \tq^{-1} Q_{a_{i-j+1}} \omega_{j|N}( \sum_{(n,m)\in \lambda^{(i-j+1)}} q_1^{m-1} \tq^n) \quad \quad \text{if} \quad j \leq i\,,\\
     Q_{a_{i-j+1}} \omega_{j|N}( \sum_{(n,m)\in \lambda^{(i-j+1)}} q_1^{m-1} \tq^n) \quad \quad \text{otherwise}\,.
\end{cases}
\end{align}
In order to understand how the computation is organised, we first consider the reorganisation of a simple subpart of the character of the tangent space to the instanton moduli space at the fixed point $\underline{\lambda}$ denoted by $\chi_{\rm inst.}\big|_{\underline{\lambda}}$:
\begin{align}
    \sum_{i=1}^N{\bf K}_i^\vee {\bf K}_i = \sum_{i=1}^N \sum_{k=1}^N \sum_{l=1}^N \frac{Q_{a_{i-l+1}}}{Q_{a_{i-k+1}}} H_{k-i,l-i}(\tq) \omega_{-k|N}(c_{\lambda^{(i-k+1)}}^\vee) \omega_{l|N}(c_{\lambda^{(i-l+1)}})\,,
\end{align}
where we used~\eqref{eq:dualpropome} and the following definitions for the extra $\tq$ factor and the so-called {\it content} $c_\lambda$ \cite{Kimura:2020jxl} of a partition $\lambda$:
\begin{align}
    H_{a,b}(\tq) := \begin{cases}
        \tq\quad \text{if} \quad (a>0 \land b \leq 0)\,,\\
        \tq^{-1} \quad \text{if} \quad (a\leq 0 \land b >0)\,,\\
        1 \quad \text{otherwise}\,,
    \end{cases} && c_\lambda := \sum_{(a,b)\in \lambda} q_1^{b-1} \tq^{a-1} \,.
\end{align}
Upon redefinitions of the summations and using \eqref{eq:prodome} we obtain:
\begin{align}
    \sum_{i=1}^N{\bf K}_i^\vee {\bf K}_i = \sum_{i=1}^N \sum_{n=1}^N \sum_{m=1}^N \frac{Q_{a_n}}{Q_{a_m}} \omega_{m-n|N}(\omega_{m-i-1|N}(c_{\lambda^{(m)}}^\vee)\cdot c_{\lambda^{(n)}})\,.
\end{align}
Finally, with~\eqref{eq:genpropome} we find:
\begin{align}\label{eq:KKcontri}
    \sum_{i=1}^N{\bf K}_i^\vee {\bf K}_i = \sum_{n=1}^N \sum_{m=1}^N \frac{Q_{a_n}}{Q_{a_m}} \omega_{m-n|N}(c_{\lambda^{(m)}}^\vee \cdot c_{\lambda^{(n)}})\,.
\end{align}
Using this result, we can read off the contribution of the untwisted sector (which arises from a term of the kind ${\bf K}_i^\vee {\bf K}_i$) to the PF:
\begin{align}
    \sum_{i=1}^N \left[-{\bf N}_i^\vee {\bf K}_i + (1-q_1^{-1}) {\bf K}^\vee_i {\bf K}_i \right] = \sum_{n,m=1}^N \frac{Q_{a_n}}{Q_{a_m}} \omega_{m-n|N} \left( (1-q_1^{-1}) c_{\lambda^{(n)}} c_{\lambda^{(m)}}^\vee - c_{\lambda^{(n)}} \right)\,.
\end{align}
From a similar reasoning, we can compute the content formula for the twisted sector (which arises from a term of the kind ${\bf K}_{i-1}^\vee {\bf K}_i$), where the extra $\tq$ factors compensate the index shifts by \eqref{eq:shiftpropome}. This contribution corresponds to:
\begin{align}
    \sum_{i=1}^N \big[-q_{1}^{-1}\tq^{-1}{\bf K}_{i-1}^\vee {\bf N}_i - &(1-q_1^{-1}) \tq^{-1} {\bf K}^\vee_{i-1} {\bf K}_i \big]\nonumber\\
    &= \sum_{n,m=1}^N \frac{Q_{a_n}}{Q_{a_m}} \omega_{m-n|N} \left(-(1-q_1^{-1}) \tq^{-1}c_{\lambda^{(n)}} c_{\lambda^{(m)}}^\vee  - q_1^{-1} q_2^{-1} c^\vee_{\lambda^{(m)}} \right)\,.
\end{align}
Finally, we see that the twisted and untwisted sector combine:
\begin{align}\label{eq:instchar}
    \chi_{\rm inst}\big|_{\underline{\lambda}} = -(1-q_3^{-1}) \sum_{n,m=1}^N \frac{Q_{a_n}}{Q_{a_m}} \omega_{m-n|N} \left(\underbrace{(1-q_1^{-1})(1-\tq^{-1}) c_{\lambda^{(n)}} c_{\lambda^{(m)}}^\vee - c_{\lambda^{(n)}}-q_1^{-1} \tq^{-1} c_{\lambda^{(m)}}^\vee}_{=f_{\lambda^{(n)}\lambda^{(m)}}(q_1,\tq)}\right)\,,
\end{align}
the factor $f_{\lambda^{(n)}\lambda^{(m)}}$ appears in the computation of the PF in the non-orbifolded background \cite{Iqbal:2003ix,Filoche:2022qxk} and $\omega_{k|N}$ allows to relate the two computations. In addition, $f_{\lambda^{(n)}\lambda^{(m)}}$ admits a representation in term of {\it arm-length} and {\it leg-length} of the partitions \cite{Iqbal:2003ix,Kimura:2020jxl}:
\begin{align}
    f_{\mu\nu}(q_1,\tq) := \sum_{\Box \in \mu} q_1^{l_{\nu}(\Box)} \tq^{-a_{\mu}(\Box)-1}+ \sum_{\Box \in \nu} q_1^{-l_{\mu}(\Box)-1} \tq^{a_{\nu}(\Box)}\,,
\end{align}
with arm-length and leg-length defined as\footnote{Partitions are understood as a collection of a finite number of non-zero entries with infinitely many zeros.}:
\begin{align}
    &a_{\mu}(\Box)=\mu_i -j \,, 
    &&l_{\mu}(\Box)=\mu_j^T -i \,, && \text{for} && \Box=(i,j)\,.
\end{align}
The relevant function that enters the orbifold computation is therefore defined by:
\begin{align}
    f_{\mu\nu}^{(p|N)}(q_1, \tq) &:= \omega_{p|N}(f_{\mu\nu}(q_1,\tq)) \nonumber\\
    &= \sum_{\tiny \begin{matrix}
        \Box \in \mu \\
        l_\mu(\Box) \equiv p \!\mod N
    \end{matrix}} q_1^{-a_\nu(\Box)-1} \tq^{l_\mu(\Box)} + \sum_{\tiny \begin{matrix}
        \Box \in \nu \\
        l_\nu(\Box) \equiv -p-1 \!\mod N
    \end{matrix}} q_1^{a_\mu(\Box)} \tq^{-l_\nu(\Box)-1}\,.
\end{align}
We define the full non-perturbative defect partition function as:
\begin{align}
    \mathcal{Z}_{\rm def.}^{\rm inst.,(N)} \left(\underline{\frak q},\mathfrak{q};\underline{Q_a},\Qr;\Qs,q_1,\tq\right) = \sum_{k_1,\ldots,k_N=0}^\infty \prod_{i=1}^N \mathfrak{q}_i^{k_i} \mathcal{Z}_{\rm def.}^{\underline{k},N}\,.
\end{align}
Taking the $6$d index~\eqref{eq:6dindex} of~\eqref{eq:instchar} and using the parametrisation~\eqref{eq:parameps} for $\varepsilon_2 \to \varepsilon_2/N$, we obtain the following expression for the non-perturbative sector of the defect PF:
\begin{tcolorbox}[colback=black!10!white,colframe=black!95!green]
\begin{align}\label{eq:defZ}
    \mathcal{Z}_{\rm def.}^{\rm inst.,(N)} \left(\underline{\frak q},\mathfrak{q};\underline{Q_a},\Qr;\Qs,q_1,\tq\right) = \sum_{\underline{\lambda}\in \mathcal{P}^N} \prod_{i=1}^N \mathfrak{q}_i^{k_i(\underline{\lambda})} \prod_{1\leq i,j\leq N} \frac{\mathcal{N}_{\lambda^{(i)}\lambda^{(j)}}^{(j-i|N)}(\Qs Q_{a_j}/Q_{a_i},\Qr;q_{1},\tq)}{\mathcal{N}_{\lambda^{(i)}\lambda^{(j)}}^{(j-i|N)}(Q_{a_j}/Q_{a_i},\Qr;q_{1},\tq)},
\end{align}
\end{tcolorbox}
\noindent
where $\mathcal{P}$ is the set of all partitions. Furthermore, we introduce the set of exponentiated gauge couplings $\underline{\mathfrak{q}}=\{\mathfrak{q}_i\}_{i\in\{1,\ldots,N\}}$ associated to each $U(1)$ gauge group (produced by the defect):
\begin{align}
    \mathfrak{q}:=\exp(2i\pi \tau) = \prod_{i=1}^N \mathfrak{q}_i\,, && \text{and} && \mathfrak{q}_j = \exp \left(2i\pi \tau_j \right)\,,
\end{align}
with $\mathfrak{q}$ the exponentiated gauge coupling of the full $U(N)$ gauge group. We recall that the charge of each $U(1)$  produced by the breaking pattern~\eqref{eq:breakingpattern} is related to the set of partitions $\underline{\lambda}$ by the colouring functions~\eqref{eq:colorf}. In addition, we define $\mathcal{N}^{(p|N)}_{\mu\nu}$ as the fractional version of the Nekrasov subfunctions which appear in the expression of the instanton PF on the non-orbifolded background \cite{Nekrasov:2002qd}:
\begin{align}\label{eq:Np}
    \mathcal{N}^{(p|N)}_{\mu\nu}(x,\Qr;q_1,\tq) :=\! \prod_{\tiny \begin{matrix}
        \Box \in \mu \\ l_\mu(\Box)\equiv p\! \mod N
    \end{matrix}}& \vartheta(x q_1^{-a_\nu(\Box)-1} q_2^{l_\mu(\Box)};\Qr)\nonumber\\
    &\times\prod_{\tiny \begin{matrix}
        \Box \in \nu \\ l_\nu(\Box)\equiv -p-1\! \mod N
    \end{matrix}} \vartheta(x q_1^{a_\mu(\Box)} q_2^{-l_\nu(\Box)-1};\Qr)\,.
\end{align}
In the following, we use the shorthand notation $\mathcal{N}^{(p|N)}(x):=\mathcal{N}^{(p|N)}_{\mu\nu}(x,\Qr;q_1,\tq)$. The fractional Nekrasov subfunctions are related to the Nekrasov subfunctions by:
\begin{align}\label{eq:prodNp}
    \mathcal{N}_{\mu\nu}(x,\Qr;q_1,\tq) = \prod_{p=1}^N \mathcal{N}^{(p|N)}(x,\Qr;q_1,\tq)\,.
\end{align}
\paragraph{}
In~\cite{shiraishi2019affinescreeningoperatorsaffine}, an analogous result to~\eqref{eq:defZ} in the $5$d setup has been obtained using an affine screening operator approach. The result of \cite{shiraishi2019affinescreeningoperatorsaffine} can be obtained from~\eqref{eq:defZ} by considering the limit $\Qr \to 0$ or by taking the $5$d index~\eqref{eq:6dindex} of~\eqref{eq:instchar}. We remark that~\eqref{eq:Np} corresponds in the $5$d limit to an equivalent representation of the fractional Nekrasov subfunctions defined in (1.1) of \cite{shiraishi2019affinescreeningoperatorsaffine}.

\subsection{Perturbative contribution}

In this subsection, we briefly discuss the perturbative contribution to the defect PF. At the level of the character computation, the contribution arises from the $6$d index of $\chi_{\rm pert.}$ as defined by~\eqref{eq:chidef}. Using the following decomposition:
\begin{align}\label{eq:expansionP2}
    {\bf P}_2^{-1} = (1-\tq \mathcal{R}_1)^{-1} = \sum_{n=0}^\infty \tq^n \mathcal{R}_1^n = \sum_{i=1}^{N} \frac{\tq^i}{1- \tq^N} \mathcal{R}_i\,,
\end{align}
we obtain:
\begin{align}
    \chi_{\rm pert.} = (1-Q_S) \sum_{i=1}^N \sum_{j=1}^N {\bf N}_i^\vee \frac{1}{1-q_1} \frac{\tq^{j}}{1-\tq^N} {\bf N}_{i+j}\,.
\end{align}
With the $6$d index of $\chi_{\rm pert.}$ we find that the perturbative contribution to the defect partition function is:
\begin{align}
    \mathcal{Z}_{\rm def.}^{\rm pert.,(N)}(\underline{Q_a},\Qr;\Qs,q_1,\tq) = \prod_{1\leq i < j \leq N} \frac{\vartheta(\frac{Q_{a_j}}{Q_{a_i}} \Qs \tq^{j-i};q_1,\tq^N)_\infty}{\vartheta(\frac{Q_{a_j}}{Q_{a_i}}  \tq^{j-i};q_1,\tq^N)_\infty} \prod_{1\leq j \leq i \leq N} \frac{\vartheta(\frac{Q_{a_i}}{Q_{a_j}} \Qs \tq^{N-j+i};q_1,\tq^N)_\infty}{\vartheta(\frac{Q_{a_i}}{Q_{a_j}}  \tq^{N-j+i};q_1,\tq^N)_\infty}\,,
\end{align}
where $\vartheta(\cdot;\cdot,\cdot)_\infty$ (defined in~\eqref{eq:appellpochdef}) is the elliptic Pochhammer symbol. This contribution can also be reformulated using double elliptic gamma functions~\cite{Narukawa:2003ltf}. The full PF in presence of a codimension-2 surface defect is therefore given by:
\begin{align}
    \mathcal{Z}_{\rm def.,(N)}\left(\underline{\frak q},\mathfrak{q};\underline{Q_a},\Qr;\Qs,q_1,\tq\right)= \mathcal{Z}_{\rm def.}^{\rm pert.,(N)}(\underline{Q_a},\Qr;\Qs,q_1,\tq) \cdot \mathcal{Z}_{\rm def.}^{\rm inst.,(N}\left(\underline{\frak q},\mathfrak{q};\underline{Q_a},\Qr;\Qs,q_1,\tq\right).
\end{align}

\subsection{Non-perturbative symmetries}\label{subsec:npsymNM1}
In this subsection we analyse some key properties of the PF~\eqref{eq:defZ}. In particular, we investigate so-called {\it non-perturbative symmetries}. These symmetries have been constructed for the instanton partition function of $A$-type LSTs without defect in \cite{Bastian:2018jlf,Filoche:2023yfm} and can be defined as a linear action on the moduli of the theory:
\begin{align}
    &\mathfrak{s}:(\{a_i\}_{i\in\{1\ldots N-1\}},\rho,S,\{\tau_i\}_{i\in \{1\ldots N\}})^T \longrightarrow \mathcal{S}\cdot(\{a_i\}_{i\in\{1\ldots N-1\}},\rho,S,\{\tau_i\}_{i\in \{1\ldots N\}})^T\,,\nonumber\\
    & \mathcal{S} \in \mathbb{M}_{2N+1,2N+1}(\mathbb{Q})\,,
\end{align}
which leaves the PF invariant. In the following, we argue that the symmetries uncovered in \cite{Bastian:2018jlf,Filoche:2023yfm} for the LST PF can be naturally generalised to symmetries of the defect PF. In the classification of \cite{Bastian:2018jlf,Filoche:2023yfm}, the symmetries are constructed by analysing the extended Kähler cone of a toric Calabi-Yau threefold \cite{Kanazawa:2016tnt} that appears in a dual F-theory construction of the LSTs. Even though there exists no similar description for the moduli space in presence of a defect, we can use the expression of the defect PF~\eqref{eq:defZ} to verify the symmetries found in \cite{Bastian:2018jlf,Filoche:2023yfm}:

\begin{itemize}
    \item The defect instanton partition function is symmetric under cyclic permutation of the Coulomb branch moduli: $Q_{a_i} \longrightarrow Q_{a_{i+1}}$, with the cyclic identification: $a_i=a_{i+N}$, $\forall i \in \{1,\ldots,N\}$. 
    \item Another manifest symmetry is the cyclic permutation of the fractional couplings: $\mathfrak{q}_i \longrightarrow \mathfrak{q}_{i+1}$\,, with the identification $\mathfrak{q}_{i+N}=\mathfrak{q}_i$, $\forall i \in \{1,\ldots,N\}$. 
    \item It is also possible to change the ordering of both the fractional couplings and Coulomb branch moduli at the same time: $\mathfrak{q}_i \longrightarrow \mathfrak{q}_{N-i}$, and $a_i \longrightarrow a_{N-i}$, $\forall i \in \{1,\ldots,N\}$.
    \item The defect instanton PF is invariant under $S \to -S +\varepsilon_1+\varepsilon_2/N$ which corresponds to a flop transition of all $(1,1)$-branes in \cite{Filoche:2023yfm} and to the Poincaré duality of~\cite{shiraishi2019affinescreeningoperatorsaffine}. This symmetry can be understood from the geometrical setup. From the $\mathbb{C}_1 \times \mathbb{C}_2/\mathbb{Z}_N \times \mathbb{T}^2$ space-time, the $\mathbb{C}_3$ and $\mathbb{C}_4$ planes play symmetrical roles, in particular the defect instanton PF should not depend on the choice between $\mathbb{C}_3/\mathbb{Z}_N \times \mathbb{C}_4$ and $\mathbb{C}_3\times \mathbb{C}_4/\mathbb{Z}_N$ and be symmetrical under the exchange $\varepsilon_3 \longleftrightarrow \varepsilon_4$, this can be tested explicitly from~\eqref{eq:generalisedvector}. Using~\eqref{eq:parameps}, this symmetry directly translates into the invariance of the defect instanton PF under $S \to -S +\varepsilon_1+\varepsilon_2/N$.
    \item We now consider the transformation $S \to S - \rho$, which leads to a symmetry of the defect instanton PF as a consequence of the quasi-periodicity property of the $\vartheta$ theta functions~\eqref{eq:appquasitheta}. In order to see this, we decompose the standard contribution of a fixed instanton contribution to the PF:
\begin{align}
    \prod_{1\leq i,j\leq N} \mathcal{N}^{(j-i|N)}_{\lambda^{(i)}\lambda^{(j)}}(\Qs Q_{a_j}/Q_{a_i}) = & \left[\prod_{1\leq i<j\leq N} \mathcal{N}^{(j-i|N)}_{\lambda^{(i)}\lambda^{(j)}}(\Qs Q_{a_j}/Q_{a_i})\right] \left[\prod_{i=1}^N \mathcal{N}_{\lambda^{(i)}\lambda^{(i)}}^{(0|N)}(\Qs)\right]\nonumber\\
    &\quad\times  \left[\prod_{1\leq i<j\leq N} \mathcal{N}^{(i-j|N)}_{\lambda^{(j)}\lambda^{(i)}}(\Qs Q_{a_i}/Q_{a_j})\right] \,.
\end{align}
Using~\eqref{eq:appquasitheta} and~\eqref{eq:Np} we have:
\begin{align}\label{eq:quasiprodNp}
    \prod_{1\leq i,j\leq N} \mathcal{N}^{(j-i|N)}_{\lambda^{(i)}\lambda^{(j)}}(\Qs \Qr^{-1} Q_{a_j}/Q_{a_i}) = \Qr^{-2|\underline{\lambda}|} \Qs^{2|\underline{\lambda}|} \prod_{1\leq i,j\leq N} \mathcal{N}^{(j-i|N)}_{\lambda^{(i)}\lambda^{(j)}}(\Qs Q_{a_j}/Q_{a_i})\,.
\end{align}
Such contributions can be systematically reabsorbed by the fractional coupling $\{\mathfrak{q}_i\}_{i\in\{1,\ldots,N\}}$ by $\mathfrak{q}_i \longrightarrow \Qr^2 \Qs^{-2} \mathfrak{q}_i$, $\forall i \in \{1,\ldots,N\}$.\footnote{We remark that this result is slightly different from the one obtained in~\cite{Filoche:2023yfm}, this only arise from the choice of representation of the theta functions~\eqref{eq:appcurlythetajacobi} between $\vartheta$ and $\theta_1$.}
    \item One can also consider the dual transformation $S \to S - \tau$. We conjecture that such a transformation can be made into a symmetry by shifting $\rho$ accordingly. We motivate this by the duality conjecture of~\cite{shiraishi2019affinescreeningoperatorsaffine,Awata:2020yxf} on the full defect PF:
    \begin{align}
        \mathcal{Z}_{\rm def.}(\underline{\mathfrak{q}},\mathfrak{q}; \underline{Q_a},\Qr; \Qs,q_1,\tq) = \mathcal{Z}_{\rm def.}(\underline{Q_a},\Qr; \underline{\mathfrak{q}},\mathfrak{q}; \Qs,q_1,\tq).
    \end{align}
    Assuming this duality, the symmetry $S \to S-\tau$ would directly following from the dual version of~\eqref{eq:quasiprodNp}.
\end{itemize}

\subsection{Particular case $N=2$}

In this subsection, we analyse the first non-trivial theory corresponding to the $U(2)$ quiver gauge theory in the presence of a defect. In particular, we will discuss two limits of interest:
\begin{itemize}
    \item The Nekrasov-Shatashvili (NS) limit \cite{Nekrasov:2009rc}: $\varepsilon_2 \to 0$. In this limit, the instanton PF is singular since $\varepsilon_2$ plays the role of a regularisation parameter for the instanton contribution. We show that the poles in $\varepsilon_2$ can be systematically cancelled by a prefactor and the leftover contribution is conjectured to be an eigenfunction of the DELL Hamiltonians \cite{Bullimore:2014awa,Koroteev:2019gqi}. This limit can be interpreted as the stationary limit from the integrable system point of view.
    \item The bulk decoupling limit: $\mathfrak{q} \to 0$. In this limit, the instanton PF reduces to a vortex PF associated with some quiver gauge theory on $\mathbb{C}\times \mathbb{T}^2$ supported on the defect. 
\end{itemize}
The non-perturbative sector of the defect PF for $N=2$ can be written in the following form:
\begin{align}\label{eq:Zn2}
    \mathcal{Z}_{\rm def.}^{\rm inst.,(2)} = \sum_{\lambda^{(1)},\lambda^{(2)}\in \mathcal{P}} \mathfrak{q}_1^{k_1(\underline{\lambda})}\mathfrak{q}_2^{k_2(\underline{\lambda})} \Bigg[\prod_{i=1}^2 \frac{\mathcal{N}_{\lambda^{(i)}\lambda^{(i)}}^{(0|2)}(\Qs)}{\mathcal{N}_{\lambda^{(i)}\lambda^{(i)}}^{(0|2)}(1)} \Bigg]  \frac{\mathcal{N}_{\lambda^{(1)}\lambda^{(2)}}^{(1|2)}(\Qs Q_a) \mathcal{N}_{\lambda^{(2)}\lambda^{(1)}}^{(1|2)}(\Qs Q_a^{-1})}{\mathcal{N}_{\lambda^{(1)}\lambda^{(2)}}^{(1|2)}(Q_a)\mathcal{N}_{\lambda^{(2)}\lambda^{(1)}}^{(1|2)}(Q_a^{-1})} \,,
\end{align}
where we used the notation: $Q_a := Q_{a_2}/Q_{a_1}$. For later convenience we define the notation:
\begin{align}
    \mathcal{Z}_{\rm def.}^{\rm inst.,(2)} =: \sum_{\lambda^{(1)},\lambda^{(2)}\in \mathcal{P}} \mathfrak{q}_1^{k_1(\underline{\lambda})}\mathfrak{q}_2^{k_2(\underline{\lambda})} Z[\lambda^{(1)},\lambda^{(2)}]\,.
\end{align}

\subsubsection{Pole subtraction and NS limit}\label{subsubsec:NSN2}

In \cite{Lee:2020hfu}, it has been argued that in the NS limit, the defect instanton PF factorises into a divergent contribution and a regular one. Using the combinatoric form of the PF, we will show how this factorisation occurs and that such factorization is non-unique. In the literature, two different prescription have been explored:
\begin{itemize}
    \item The {\it bulk normalisation} procedure \cite{Lee:2020hfu,Koroteev:2019gqi,Kimura:2022zsx}: in this prescription, the defect PF is divided by the bulk PF, i.e. the PF obtained from the non-orbifolded background. We will show that:
    \begin{align}\label{eq:regbulk}
        \Psi^{(2)}:=\frac{\mathcal{Z}_{\rm def.}^{\rm inst.,(2)}}{\mathcal{Z}_{\rm bulk}^{\rm inst.,(2)}} = \mathcal{O}(\varepsilon_2^0)\,.
    \end{align}
    It is useful to notice that the bulk PF can be obtained from the defect PF by restricting the summations over all partitions in~\eqref{eq:Zn2} to summations over the following set of {\it bulk partitions}:
    \begin{align}\label{eq:bulkPF}
        \mathcal{Z}_{\rm bulk}^{\rm inst.,(2)} = \sum_{\lambda^{(1)},\lambda^{(2)}\in \mathcal{P}_{\rm bulk}^{(2)}} \mathfrak{q}_1^{k_1(\underline{\lambda})}\mathfrak{q}_2^{k_2(\underline{\lambda})} Z[\lambda_1,\lambda_2]\,,
    \end{align}
    where the set of bulk partitions is defined as:
    \begin{align}
        \mathcal{P}_{\rm bulk}^{(N=2)} := \{\lambda = (\lambda_1 \ldots \lambda_l) \in \mathcal{P} \quad \text{such that}  \quad \lambda_i^T = 0\!\mod 2\,, \,\,\, \forall i \in \{1\ldots \lambda_1\} \}\,.
    \end{align}
    Such restriction on the set of all partitions can be understood graphically as it allows to group boxes by two vertically:
    \begin{align}
    \begin{ytableau}
        0&1\\
        0&1\\
        2\\
        2
    \end{ytableau} \to \begin{ytableau}
        0&1\\2
    \end{ytableau}\,, && \begin{ytableau}
        0 & 1 & 2 & 3 \\0 & 1 & 2 & 3
    \end{ytableau} \to \begin{ytableau}
        0 & 1 & 2 & 3 
    \end{ytableau}\,,
    \end{align}
    where boxes grouped together are labelled by the same integer.
    \item The {\it Shiraishi normalisation} procedure \cite{shiraishi2019affinescreeningoperatorsaffine,Mironov:2023rzb}: the defect PF is divided by the following normalisation function $\mathfrak{Z}^{(2)}$:
    \begin{align}\label{eq:curlyZ}
        \mathfrak{Z}^{(2)} = \sum_{\tiny \begin{matrix}
            \lambda^{(1)},\lambda^{(2)}\in \mathcal{P}\\ k_1(\underline{\lambda})=k_2(\underline{\lambda})
        \end{matrix}} \mathfrak{q}_1^{k_1(\underline{\lambda})}\mathfrak{q}_2^{k_2(\underline{\lambda})} Z[\lambda^{(1)},\lambda^{(2)}] \,,
    \end{align}
    we will show that:
    \begin{align}\label{eq:shiraishireg}
        \frac{\mathfrak{Z}^{(2)}}{\mathcal{Z}_{\rm bulk}^{\rm inst.,(2)}}=\mathcal{O}(\varepsilon_2^0)&& \text{and}&&\widetilde{\Psi}^{(2)}:=\frac{\mathcal{Z}_{\rm def.}^{\rm inst.,(2)}}{\mathfrak{Z}^{(2)}}=\mathcal{O}(\varepsilon_2^0)\,.
    \end{align}
\end{itemize}
Singular contributions in the NS limit, arise from particular partitions configuration. Since $Q_a \neq 1$, such configurations are characterised by the following condition:
\begin{align}\label{eq:condNS}
    \mathcal{N}^{(0|2)}_{\lambda^{(i)}\lambda^{(i)}}(1) = \mathcal{O}(\varepsilon_2^{n})\,, \quad n\geq1 \,,&& \Leftrightarrow&& \exists \Box \in \lambda^{(i)}\,, \quad l_{\lambda^{(i)}}(\Box) = 0 \! \mod 2\,, \quad a_{\lambda^{(i)}}(\Box)=0\,,
\end{align}
for instance for the following partitions:
\begin{align}
    \ydiagram[*(white) \bullet]
    {4+0,1+1,2+0,0+1,1+0}
    *[]{4,2,2,1,1} && \ydiagram[*(white) \bullet]
    {0+1,1+0}
    *[]{1,1} && \ydiagram[*(white) \bullet]
    {4+0,3+0,1+0}
    *[]{4,3,1}
\end{align}
boxes that satisfy the condition on the r.h.s. of~\eqref{eq:condNS} are denoted $\tiny{\ydiagram[*(white) \bullet]{0+1}*[]{1}}$. If one consider a general contribution $Z[\lambda^{(1)},\lambda^{(2)}]$, this term has a pole of order $n\geq 0$ in the NS limit where $n$ is the number of boxes in $\lambda^{(1)}$ and $\lambda^{(2)}$ satisfying the condition on the r.h.s. of~\eqref{eq:condNS}. The proof of regularity of the normalisation procedures will rely on a systematic subtraction of these poles.

\paragraph{}
In order to understand the pole subtraction, we start by giving some definitions. We define a fusion operation on partitions $\forall \lambda=(\lambda_1,\ldots,\lambda_l) \in \mathcal{P}\,, \quad \forall \mu=(\mu_1,\ldots,\mu_k) \in \mathcal{P}$:
\begin{align}\label{eq;plusoperation}
\lambda\oplus\mu:=((\lambda\oplus\mu)_1 ,\ldots,(\lambda\oplus\mu)_{l+k})\,, \quad \text{with} \quad \begin{cases} (\lambda\oplus\mu)_i \in \{\lambda_1,\ldots,\lambda_l,\mu_1, \ldots, \mu_k\}\,,\\
(\lambda \oplus\mu)_1 \geq \ldots \geq (\lambda \oplus \mu)_{l+k}\,.\end{cases}
\end{align}
for example we have:
\begin{align}
    (5,1)\oplus(6,2) = (6,5,2,1)\,, &&(3,2,2,1)\oplus(4,2,1) = (4,3,2,2,2,1,1)\,.
\end{align}
We also define the set of all sub bulk partitions of a given partitions, $\forall \mu \in \mathcal{P}$:
\begin{align}
    \mathcal{S}^{(2)}(\mu):=\{\mu^{(b)}\in \mathcal{P}_{\rm bulk}^{(2)} \quad \text{such that}\quad \exists \lambda \in \mathcal{P} \quad \text{that satisfies} \quad \lambda \oplus \mu^{(b)}=\mu \}\,,  
\end{align}
for instance we have:
\begin{align}
    \mathcal{S}^{(2)}\big((3,3,2,1,1,1,1) \big)= \{\emptyset,(3,3,1,1,1,1),(3,3,1,1),(3,3),(1,1,1,1),(1,1)\}\,.
\end{align}
We define a subtraction on the set of partitions:
\begin{align}\label{eq:minusoperation}
    \forall \lambda,\mu \in \mathcal{P}\,, \quad \text{such that}\quad \mu \subset\lambda, \quad \lambda \ominus\mu:=\lambda\setminus\mu\,, 
\end{align}
where the last operation is understood as an operation on sets of integers. For example:
\begin{align}
    (3,2,2,1)\ominus(2,2) = (3,1)\,, && (4,4,3,2,1,1)\ominus(3) = (4,4,2,1,1)\,.
\end{align}
These definitions allow to formulate a recursive relation on $Z[\mu,\nu]$ that guarantees a systematic cancellation of poles in the NS limit:
\begin{tcolorbox}[colback=black!10!white,colframe=black!95!green]
\begin{align}\label{eq:bilin1}
    Z^{(0)}[\mu,\nu] := Z[\mu,\nu] - \sum_{\tiny{\begin{matrix}\alpha \in \mathcal{S}^{(2)}(\mu)\\ \beta \in \mathcal{S}^{(2)}(\nu)\\(\alpha,\beta)\neq(\emptyset,\emptyset)\end{matrix}}} Z[\alpha,\beta]\cdot Z^{(0)}[\mu \ominus \alpha,\nu \ominus \beta] = \mathcal{O}(\varepsilon_2^0)\,,
\end{align}
\end{tcolorbox}
\noindent
this relation is for now conjectured but has been tested explicitly using the combinatoric expression of $Z[\mu,\nu]$ for all $\mu,\nu \in \mathcal{P}$ such that $|\mu|+|\nu|\leq 12$. We remark that~\eqref{eq:bilin1} is initialised by a set of partitions which we will call {\it seed partitions} defined in the following manner:
\begin{align}
    \mathcal{P}_{\rm seed}^{(2)} := \{ \lambda=(\lambda_1,\ldots,\lambda_l)\in \mathcal{P} \quad \text{such that} \quad \lambda_{i+1} < \lambda_{i}\,,\quad i \in\{1,\ldots, l-1\}\}\,,
\end{align}
such restriction and its generalisation is known as the Burge condition~\cite{BURGE1993210} and have been discussed in several other contexts such as minimal models \cite{Bershtein:2014qma,Alkalaev:2014sma}. We give some examples of seed partitions:
\begin{align}
    \ydiagram{1}\,, && \ydiagram{3,2}\,, && \ydiagram{4,2,1}\,,
\end{align}
using the criterion~\eqref{eq:condNS} we can directly see that for $\mu,\nu \in \mathcal{P}_{\rm seed}^{(2)}$, we have $Z[\mu,\nu]=\mathcal{O}(\varepsilon_2^0)$, in addition for $\lambda \in \mathcal{P}_{\rm seed}^{(2)}$, we have $\mathcal{S}^{(2)}(\lambda)=\{\emptyset\}$. This gives the initialisation of~\eqref{eq:bilin1}: $Z^{(0)}[\mu,\nu]=Z[\mu,\nu]=\mathcal{O}(\varepsilon_2^0)$. Furthermore, the set of seed partitions provides a natural basis on which the defect instanton PF can be decomposed:
\begin{align}
    \mathcal{Z}_{\rm def.}^{\rm inst.,(2)} = \sum_{\mu,\nu \in \mathcal{P}^{(2)}_{\rm seed}} \mathfrak{q}_1^{k_1(\mu,\nu)} \mathfrak{q}_2^{k_2(\mu,\nu)} \sum_{\alpha,\beta \in \mathcal{P}_{\rm bulk}^{(2)}} \mathfrak{q}^{\frac{|\alpha|+|\beta|}{2}}Z[\mu \oplus \alpha,\nu \oplus \beta]\,.
\end{align}
Such a decomposition relies on a decomposition property of the set of partitions, $\forall \lambda^{(s)} \in \mathcal{P}_{\rm seed}^{(2)}$ we define:
\begin{align}
    \mathcal{T}^{(2)}(\lambda^{(s)}) := \left\{ \lambda^{(s)}\oplus\lambda^{(b)}\,, \quad \forall \lambda^{(b)} \in \mathcal{P}_{\rm bulk}^{(2)} \right\}\,,
\end{align}
and we have that:
\begin{align}
    \mathcal{P} = \bigcup_{\mu \in \mathcal{P}_{\rm seed}^{(2)}} \mathcal{T}^{(2)}(\mu)\,, \quad \text{and} \quad \mathcal{T}^{(2)}(\mu) \,\cap \, \mathcal{T}^{(2)}(\nu)=\emptyset\,, \quad \forall \mu\neq \nu \in \mathcal{P}_{\rm seed}^{(2)}\,.
\end{align}
\paragraph{}
We now come to the proof of regularity of the bulk normalisation procedure~\eqref{eq:regbulk}. More precisely, we will show that this regularity is guaranteed by~\eqref{eq:bilin1}. We start by defining $\forall \mu,\nu \in\mathcal{P}_{\rm seed}^{(2)}$:
\begin{align}
    Z[(\mu,\nu)\oplus n] := \sum_{\tiny \begin{matrix}
        \alpha,\beta \in \mathcal{P}_{\rm bulk}^{(2)} \\ |\alpha|+|\beta|=2n
    \end{matrix}} Z[\mu \oplus \alpha,\nu \oplus\beta]\,,
\end{align}
this quantity corresponds to all possible ways to add bulk partitions to the seed partitions $\mu,\nu$ such that $2n$ boxes have been added. We notice that using this notation the bulk instanton PF~\eqref{eq:bulkPF} is:
\begin{align}
    \mathcal{Z}_{\rm bulk}^{\rm inst.,(2)} = \sum_{n\geq 0} \mathfrak{q}^n Z[(\emptyset,\emptyset)\oplus n]\,, \quad Z_n^{(2)} := Z[(\emptyset,\emptyset)\oplus n]\,.
\end{align}
Using~\eqref{eq:bilin1}, we can write a recursive relation on $Z^{(0)}[(\mu,\nu) \oplus n]$, in particular we can use~\eqref{eq:bilin1} $n$ times to write:
\begin{align}\label{eq:Z0pn}
    Z^{(0)}[(&\mu,\nu) \oplus n] = Z[(\mu,\nu) \oplus n] - Z_1^{(2)} Z[(\mu,\nu)\oplus(n-1)] - (Z_2^{(2)}-(Z_1^{(2)})^2)Z[(\mu,\nu)\oplus(n-2)]\nonumber\\
    &- \ldots - P_k(-Z_1^{(2)},\ldots,-Z_k^{(2)}) Z[(\mu,\nu) \oplus (n-k)]-\ldots -P_n(-Z_1^{(2)},\ldots,-Z_n^{(2)})Z[\mu,\nu]\,,
\end{align}
where $P_n$ is a $n$-variable polynomial. A similar relation can be obtained for $Z^{(0)}[(\mu,\nu)\oplus (n+1)]$, using~\eqref{eq:Z0pn} we deduce a recursion relation on $P_n$ and obtain with $P_0 =1$:
\begin{align}
    P_{n+1}(-Z_1^{(2)},\ldots,-Z_n^{(2)},-Z_{n+1}^{(2)}) = \sum_{k=0}^n \binom{n}{k} (-Z_{k+1}^{(2)}) P_{n-k}(-Z_1^{(2)},\ldots,-Z_{n-k}^{(2)})\,,
\end{align}
this relation is the recursive definition of the complete Bell polynomials which can be seen as the series expansion of the bulk normaliser:
\begin{align}
    \sum_{n=0}^\infty P_n(-Z_1^{(2)},\ldots,-Z_n^{(2)}) \mathfrak{q}^n = \frac{1}{1+Z_1^{(2)} \mathfrak{q}+Z_2^{(2)} \mathfrak{q}^2 + \ldots + Z_n^{(2)} \mathfrak{q}^n + \ldots} = \frac{1}{\mathcal{Z}_{\rm bulk}^{\rm inst.,(2)}}\,,
\end{align}
putting all the pieces together we obtain that:
\begin{align}\label{eq:regrelbulk}
    \Psi^{(2)} = \sum_{\mu,\nu \in \mathcal{P}_{\rm seed}^{(2)}} \mathfrak{q}_1^{k_1(\mu,\nu)} \mathfrak{q}_2^{k_2(\mu,\nu)} \sum_{n=0}^\infty \mathfrak{q}^n Z^{(0)}[(\mu,\nu) \oplus n]=\mathcal{O}(\varepsilon_2^0)\,.
\end{align}

\paragraph{}
The regularity of Shiraishi's normalisation~\eqref{eq:shiraishireg} in the NS limit follows from a similar argument. We notice that the normaliser defined by~\eqref{eq:curlyZ} can be decomposed in the following way:
\begin{align}
    \mathfrak{Z}^{(2)} = \mathcal{Z}_{\rm bulk}^{\rm inst.,(2)}+\sum_{\tiny \begin{matrix}
        \mu,\nu\in \mathcal{P}_{\rm seed}^{(2)}\\k_1(\mu,\nu)=k_2(\mu,\nu)
    \end{matrix}} \mathfrak{q}_1^{k_1(\mu,\nu)} \mathfrak{q}_2^{k_2(\mu,\nu)}\sum_{n=0}^\infty Z[(\mu,\nu) \oplus n]\,,
\end{align}
using this formulation, we directly see that the relation~\eqref{eq:Z0pn} guarantees the regularity of $\mathfrak{Z}^{(2)}/\mathcal{Z}_{\rm bulk}^{\rm inst.,(2)}$ and~\eqref{eq:regrelbulk} gives the regularity of Shiraishi's normalisation~\eqref{eq:shiraishireg}.

\paragraph{}
Similarly any normaliser that include the bulk PF and complete summations over $\mathcal{T}^{(2)}(\mu)$ for $\mu \in \mathcal{P}_{\rm seed}^{(2)}$ will regularize the defect instanton PF. Among all possible regulators which can be engineered using this principle the two possibilities discussed previously stand apart: $\mathcal{Z}_{\rm bulk}^{\rm inst.,(2)}$ is the minimal regulator while $\mathfrak{Z}^{(2)}$ systematically suppress contribution such that $k_1(\underline{\lambda})=k_2(\underline{\lambda})$, from the integrable system perspective this could correspond to contribution that only depend on the center of mass coordinate. Another way of understanding this freedom of choice for the regulator is to use~\eqref{eq:prodNp} to reformulate the defect instanton PF in the following way:
\begin{align}
    \mathcal{Z}_{\rm def.}^{\rm inst.,(2)} = \sum_{\underline{\lambda}} \mathfrak{q}^{|\underline{\lambda}|} \mathcal{Z}_{\rm bulk}^{\rm inst.^{(2)}}\big|_{\underline{\lambda},\varepsilon_2 \to \frac{\varepsilon_2}{2}} \underbrace{\prod_{i=1}^2 \mathfrak{q}_i^{k_i - |\underline{\lambda}|} \frac{\mathcal{N}_{\lambda^{(i)}\lambda^{(i)}}^{(1|2)}(1)}{\mathcal{N}_{\lambda^{(i)}\lambda^{(i)}}^{(1|2)}(\Qs)}  \frac{\mathcal{N}_{\lambda^{(1)}\lambda^{(2)}}^{(0|2)}(Q_a)\mathcal{N}_{\lambda^{(2)}\lambda^{(1)}}^{(0|2)}(Q_a^{-1})}{\mathcal{N}_{\lambda^{(1)}\lambda^{(2)}}^{(0|2)}(\Qs Q_a) \mathcal{N}_{\lambda^{(2)}\lambda^{(1)}}^{(0|2)}(\Qs Q_a^{-1})}}_{=\mathcal{Z}_{\rm surf.}^{\rm inst.,(2)}\big|_{\underline{\lambda}}} \,,
\end{align}
in this formulation, any term that is singular in the NS limit is directly encapsulated in $\mathcal{Z}_{\rm bulk}^{\rm inst.,(2)}\big|_{\underline{\lambda}}$ which means the bulk instanton contribution at fixed point $\underline{\lambda}$. In \cite{Lee:2020hfu}, this principle has been used to argue that in the NS limit, the bulk prefactor locks the partitions to the limit shape configurations. However, in general because of its $\mathcal{N}^{(1|2)}(1)$ factor in the numerator, $\mathcal{Z}_{\rm surf.}^{\rm inst.,(2)}\big|_{\underline{\lambda}}$ has some zeros in the NS limit which modify the pole structure, we expect that such zeros allow some freedom in the definition of the limit shape configurations.

\subsubsection{Bulk decoupling limit}\label{Sect:BulkHamilton}
In the bulk decoupling limit, we consider the limit $\mathfrak{q}\to 0$\,. Physically, this corresponds to reducing the defect instanton PF to the vortex PF of the world-volume theory of the defect. This theory correspond to a $\mathcal{N}=2^*$ $U(1)$ quiver gauge theory with a $U(2)$ flavor group on $\mathbb{C}_{\varepsilon_1}\times \mathbb{T}^2$ \cite{Koroteev:2019byp}. At the combinatorial level, the relation $\mathfrak{q}=\mathfrak{q}_1 \mathfrak{q}_2$ enforces that $k_1(\lambda^{(1)},\lambda^{(2)})$ or $k_2(\lambda^{(1)},\lambda^{(2)})$ is fixed to $0$. We choose $k_2=0$. This choice does not modify the final result since the defect instanton PF is manifestly invariant under the exchange of the two partitions together with the exchange of couplings as discussed in subsection~\ref{subsec:npsymNM1}. The partitions are therefore given by $(\lambda^{(1)}=(n),\lambda^{(2)}=\emptyset)$ with $n \in \mathbb{N}$. Such configurations belong to a subset of $\mathcal{P}_{\rm seed}^{(2)}$. One can then safely take the NS limit $\varepsilon_2\to 0$ which leads to the following vortex PF:
\begin{align}\label{eq:vortexN2}
    \mathcal{Z}_{\mathbb{C}\times \mathbb{T}^2}^{\rm inst.} = \sum_{n=0}^\infty \mathfrak{q}_1^n \frac{\vartheta(\frac{q_1}{\Qs};q_1)_n \vartheta(\frac{Q_a}{\Qs q_1};q_1)_n}{\vartheta(q_1;q_1)_n \vartheta(\frac{Q_a}{q_1};q_1)_n}\,.
\end{align}
This result corresponds to an unbalanced version of the elliptic hypergeometric function $\,_2E_1$ defined in~\eqref{eq:appellhyp}: This is a consequence of the shift symmetry discussed in subsection~\ref{subsec:npsymNM1}. Such an elliptic hypergeometric function naturally comes with a $q$-difference equation. Upon defining:
\begin{align}
    {\bf p}_1 := q_1^{\mathfrak{q}_1 \partial_{\mathfrak{q}_1}}\,, \quad \quad \text{satisfying} \quad \quad {\bf p}_1 \mathfrak{q}_1 = q_1 \mathfrak{q}_1 {\bf p}_1\,, 
\end{align}
the operator:
\begin{align}
    {\bf H}_{\mathbb{C}\times \mathbb{T}^2} := \vartheta({\bf p}_1;\Qr) \vartheta \bigg(\frac{Q_a}{q_1} {\bf p_1;\Qr}\bigg) - \mathfrak{q}_1 \vartheta \bigg( \frac{q_1}{\Qs} {\bf p}_1 ;\Qr\bigg) \vartheta \bigg(\frac{Q_a}{\Qs q_1} {\bf p}_1 ;\Qr\bigg)\,,
\end{align}
annihilates the vortex PF~\eqref{eq:vortexN2}. The vortex PF can be further reduced by considering the Higgsing $Q_a \to q_1^{\gamma_1}\Qs$, with $\gamma_1 \in \mathbb{N}$. Such a Higgsing condition imposes an additional defect to realize the intersecting defect configuration (see, e.g.,~\cite{Jeong:2021rll}), which allows from the integrable system viewpoint to construct the wave-function of the dual elliptic-RS system~\cite{Koroteev:2019gqi}. For $\gamma_1=2$, which corresponds to the minimal case in our parametrisation, the vortex PF simplifies to:
\begin{align}
    \mathcal{Z}_{\mathbb{C}\times \mathbb{T}^2}^{{\rm inst.},\gamma_1=2} = \sum_{n=0}^\infty \mathfrak{q}_1^n \frac{\vartheta(\frac{q_1}{\Qs};q_1)_n}{\vartheta(q_1 \Qs;q_1)_n}\,,
\end{align}
and the annihilation operator is given by:
\begin{align}
    {\bf H}_{\mathbb{C}\times \mathbb{T}^2}^{\gamma_1=2} := \vartheta(q_1\Qs {\bf p}_1;\Qr) - \mathfrak{q}_1 \vartheta \bigg( \frac{q_1}{\Qs} {\bf p}_1;\Qr\bigg)\,.
\end{align}
As discussed in~\cite{Koroteev:2019gqi,Gorsky:2021wio}, such an operator annihilates the elliptic generalisation of MacDonald polynomials.
One can read off the expectation value of the Wilson surface wrapping the torus from the constant term of this elliptic difference operator. In contrast to the 5d/3d or 4d/2d system, this constant term cannot be separated from the operator acting on the partition function due the modular property specific to the 6d/4d system.

\subsection{Generalisation to arbitrary $N$}

In this subsection, we discuss the generalisation of the result for the NS limit $\varepsilon_2 \to 0$ discussed for $N=2$ in subsection~\ref{subsubsec:NSN2}. We give elements to understand how the subtraction of the poles in $\varepsilon_2$ can be performed for general $N$. We then proceed to derive the bulk decoupling limit as defined for $N=2$ and check that the obtained PF corroborate existing results and conjecture in literature. In the following, we will use the shorthand notation:
\begin{align}
Z[\lambda^{(1)},\ldots,\lambda^{(N)}]:=\prod_{1\leq i,j\leq N} \frac{\mathcal{N}_{\lambda^{(i)}\lambda^{(j)}}^{(j-i|N)}(\Qs Q_{a_j}/Q_{a_i},\Qr;q_{1},\tq)}{\mathcal{N}_{\lambda^{(i)}\lambda^{(j)}}^{(j-i|N)}(Q_{a_j}/Q_{a_i},\Qr;q_{1},\tq)}\,,
\end{align}
for the contribution to the defect instanton PF labelled by the $N$-tuple of partitions $\underline{\lambda}=(\lambda^{(1)},\ldots,\lambda^{(N)})$.

\subsubsection{Pole subtraction and NS limit}\label{subsec:NSgenN}

For general $N$, we observe similar properties for the defect instanton PF in the NS limit $\varepsilon_2 \to 0$ as for the $N=2$, i.e:
\begin{itemize}
    \item The defect instanton PF normalized by the bulk instanton PF is regular in the NS limit. The bulk partition function can be obtained as a restriction of the defect instanton PF corresponding to:
    \begin{align}
        \mathcal{Z}_{\rm bulk}^{\rm inst.,(N)}:= \sum_{\underline{\lambda} \in \big(\mathcal{P}_{\rm bulk}^{(N)}\big)^N}\prod_{i=1}^N \mathfrak{q}_i^{k_i(\underline{\lambda})} Z[\lambda^{(1)},\ldots,\lambda^{(N)}]\,,
    \end{align}
    where the set of bulk partitions is now defined as the set of partitions for which all boxes can be grouped by $N$ vertically:
    \begin{align}\label{eq:defPbulkN}
        \mathcal{P}_{\rm bulk}^{(N)}:= \left\{ \lambda = (\lambda_1,\ldots,\lambda_l) \in \mathcal{P} \,\, |\,\, \lambda_i^T = 0 \!\!\! \mod N\,, \quad \forall i \in \{1,\ldots,\lambda_1\} \right\}\,.
    \end{align}
    \item The defect instanton PF normalized by Shiraishi's normalizer factor defined by:
    \begin{align}
        \mathfrak{Z}^{(N)}:= \sum_{\tiny \begin{matrix}
            \underline{\lambda}\in \mathcal{P}^N\\
            k_1(\underline{\lambda}) = \ldots =k_N(\underline{\lambda})
        \end{matrix}} \prod_{i=1}^N \mathfrak{q}_i^{k_i(\underline{\lambda})} Z[\lambda^{(1)},\ldots,\lambda^{(N)}]\,,
    \end{align}
    is regular in the NS limit.
    \item Any normalizer which can be decomposed as the sum of the bulk instanton PF and a sum of seeds with all possible way to insert bulk partitions in them will lead to a regular NS limit:
    \begin{align}
        \mathcal{N} = \mathcal{Z}_{\rm bulk}^{\rm inst.,(N)} + \sum_{\underline{\mu} \in \mathcal{S}\subset  \big (\mathcal{\mathcal{P}}_{\rm seed}^{(N)} \big)^N} \sum_{\underline{\nu} \in \mathcal{T}^{(N)}(\underline{\mu})} \prod_{i=1}^N \mathfrak{q}_i^{k_i(\underline{\nu})} Z[\nu^{(1)},\ldots,\nu^{(N)}]\,,
    \end{align}
    is a valid normalizer in the NS limit, where $\mathcal{S}$ is any subset of $(\mathcal{\mathcal{P}}_{\rm seed}^{(N)} \big)^N$ and we have the following definitions:
    \begin{align}\label{eq:seedsN}
        \mathcal{P}_{\rm seed}^{(N)}:= \left\{\lambda=(\lambda_1,\ldots,\lambda_l) \in \mathcal{P}\,\, | \, \, \lambda_{i+N-1}< \lambda_i \,, \quad \forall i \in \{1,\ldots,l-N+1\}  \right\}\,,
    \end{align}
    and for $\underline{\mu} \in \big(\mathcal{P}_{\rm seed}^{(N)}\big)^N$:
    \begin{align}\label{eq:defTN}
        \mathcal{T}^{(N)}(\underline{\mu}):=\left\{(\mu^{(1)} \oplus \lambda^{(1)},\ldots,\mu^{(N)} \oplus \lambda^{(N)})\,, \quad\forall \underline{\lambda}=(\lambda^{(1)},\ldots,\lambda^{(N)}) \in \big(\mathcal{P}^{(N)}_{\rm bulk} \big)^N \right\}\,
    \end{align}
    where the $\oplus$ operation on partitions is defined by~\eqref{eq;plusoperation}.
\end{itemize}
Similarly to the case $N=2$, the regularity in the NS limit relies on a recursive relation which allows to systematically cancel poles in $\varepsilon_2$. We can define the set of all bulk sub-partitions of a given partition, $\forall \underline{\mu} \in \mathcal P^N$:
\begin{align}\label{eq:defSN}
    \mathcal{S}^{(N)}(\underline{\mu}):= \left\{ \underline{\nu} \in \big(\mathcal{P}_{\rm bulk}^{(N)} \big)^N \, \, | \,\, \exists \underline{\lambda}\in \mathcal{P}^N\quad \text{with}\quad \underline{\mu}=(\nu^{(1)} \oplus \lambda^{(1)},\ldots,\nu^{(N)} \oplus \lambda^{(N)}) \right\}\,,
\end{align}
the recursive relation is then given by:
\begin{tcolorbox}[colback=black!10!white,colframe=black!95!green]
\begin{align}
    Z^{(0)}[\underline{\lambda}]=Z[\underline{\lambda}]-\!\! \sum_{\tiny \begin{matrix}
        \underline{\mu} \in \mathcal{S}^{(N)}(\underline{\lambda})\\
        (\mu^{(1)},\ldots,\mu^{(N)})\neq(\emptyset,\ldots,\emptyset)
    \end{matrix}}\!\! Z[\underline{\mu}]  Z^{(0)}[\lambda^{(1)} \ominus \mu^{(1)},\ldots,\lambda^{(N)} \ominus \mu^{(N)}]\,,
\end{align}
\end{tcolorbox}
\noindent
and we observe that:
\begin{align}
    Z^{(0)}[\lambda^{(1)},\ldots,\lambda^{(N)}] = \mathcal{O}(\varepsilon_2^0)\,, \quad \quad \forall \underline{\lambda}\in \mathcal{P}^N\,.
\end{align}
One can easily see that such a recursion relation is initialized by seed partitions defined by~\eqref{eq:seedsN} which are by construction regular in the NS limit. We then conjecture that this relation translates into the following, $\forall \underline{\mu} \in \big(\mathcal{P}_{\rm seed}^{(N)} \big)^N$:
\begin{align}
    Z^{(0)}[\underline{\mu}\oplus n] = \sum_{k=0}^n P_k(-Z_1^{(N)},\ldots,-Z_k^{(N)}) Z[\underline{\mu} \oplus(n-k)]\,,
\end{align}
where we have:
\begin{align}
    Z[\underline{\mu} \oplus(n-k)] := \sum_{\tiny \begin{matrix}\underline{\nu} \in \big(\mathcal{P}_{\rm bulk}^{(N)} \big)^N \\
    |\underline{\nu}|=Nn\end{matrix}} Z[\mu^{(1)}\oplus \nu^{(1)},\ldots, \mu^{(N)} \oplus \nu^{(N)}]\,, && Z_k^{(N)}:=Z[(\underbrace{\emptyset,\ldots,\emptyset}_{N\text{ times}}) \oplus k]\,,
\end{align}
and $\{P_k\}_{k\in\mathbb{N}}$ is the set of complete Bell polynomials.

\paragraph{}
We can motivate the conjecture appearing in this subsection by the following re-writing of the defect PF:
\begin{align}
    &\mathcal{Z}_{\rm def.}^{\rm inst.,(N)} = \sum_{\underline{\lambda}} \mathfrak{q}^{|\underline{\lambda}|} \mathcal{Z}_{\rm bulk.}^{\rm inst.,(N)}\big|_{\underline{\lambda},\varepsilon_2 \to \frac{\varepsilon_2}{N}}  \prod_{i=1}^N \mathfrak{q}_i^{k_i(\underline{\lambda})-|\underline{\lambda}|}\mathcal{Z}_{\rm surf.}^{\rm inst.,(N)} \big|_{\underline{\lambda}}\,,\nonumber\\
    &\mathcal{Z}_{\rm surf.}^{\rm inst.,(N)} \big|_{\underline{\lambda}} := \prod_{1\leq i,j\leq N} \prod_{\tiny \begin{matrix}
        p=1\\
        p\neq j-i
    \end{matrix}}^N \frac{\mathcal{N}^{(p|N)}(\Qs Q_{a_j}/Q_{a_i})}{\mathcal{N}^{(p|N)}( Q_{a_j}/Q_{a_i})}\,,
\end{align}
where we observe that $\mathcal{Z}_{\rm surf.}^{\rm inst.,(N)} \big|_{\underline{\lambda}}=\mathcal{O}(\varepsilon_2^k)$ with $k \geq 0$, the pole structure is then encapsulated in $\mathcal{Z}_{\rm bulk.}^{\rm inst.,(N)} \big|_{\underline{\lambda}}$ and one can then normalize by the bulk PF. 

\subsubsection{Bulk decoupling limit}\label{sec:bulkdecouplingN}
\begin{figure}[htbp]
    \centering
    \begin{tikzpicture}
        \draw[ thick] (-2.65,-0.65) -- (-2.65,0.65) -- (-1.35,0.65) -- (-1.35,-0.65) -- (-2.65,-0.65);
        \draw[ thick] (-1.35,0) -- (0,0);
        \draw[ thick] (0,0) -- (2,0);
        \draw[ thick,dashed] (2,0) -- (4.5,0);
        \draw[ thick] (5,0) -- (7,0);
        \draw[ thick,fill=white] (0,0) circle (0.65cm);
        \draw[ thick,fill=white] (2,0) circle (0.65cm);
        \draw[ thick,fill=white] (5,0) circle (0.65cm);
        \draw[ thick,fill=white] (7,0) circle (0.65cm);
        \node at (-2,0) {$N$};
        \node at (0,0) {$N-1$};
        \node at (2,0) {$N-2$};
        \node at (5,0) {\large $2$};
        \node at (7,0) {\large $1$};
    \end{tikzpicture}
    \caption{Quiver describing the $4$d gauge theory in the defect world-volume.}
    \label{fig:quiverCT}
\end{figure}
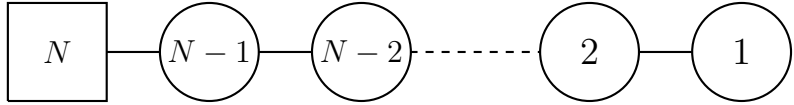

In the bulk decoupling limit $\mathfrak{q}\to0$\,, the defect instanton PF simplifies and is now interpreted as a vortex PF for the world-volume theory of the defect. This theory is a $\mathcal{N}=2$ quiver gauge theory described by the quiver of~\figref{fig:quiverCT} on a $\mathbb{C}_{\varepsilon_1}\times \mathbb{T}^2$ space-time \cite{Koroteev:2019gqi}. 
From a combinatoric point of view, the set of partitions contributing to the vortex PF $\mathcal{S}_{\rm vortex}^{(N)}$ corresponds to:
\begin{equation}
    \mathcal{S}_{\rm vortex}^{(N)}:=\left\{ \underline{\lambda}=(\lambda^{(1)},\ldots,\lambda^{(N)}) \in \mathcal{P}^N \, \, | \,\, \lambda^{(i),T}_{1} \leq N-i \right\}\,,
\end{equation}
in particular we remark that $\mathcal{S}_{\rm vortex}^{(N)}\subset \big(\mathcal{P}_{\rm seed}^{(N)} \big)^N$ which guarantees the regularity of the vortex PF in the NS limit as discussed in subsection~\ref{subsec:NSgenN}. The NS limit can be safely performed to obtain:
\begin{align}\label{eq:bulkdecouplinggeneralN}
    &\mathcal{Z}_{\mathbb{C}\times \mathbb{T}^2}^{(N)} := \sum_{\tiny \begin{matrix}n_{N-i}^{(i)}\geq\ldots\geq n_{1}^{(i)}\geq0 \\i \in \{1,\ldots,N-1\}\end{matrix}} \prod_{i=1}^{N-1}\mathfrak{q}_i^{\sum_{k=1}^i n_{i-k+1}^{(k)}} \prod_{k=1}^{N-i} \frac{\vartheta(\frac{q_1}{\Qs};q_1)_{n_{k}^{(i)}-n_{k-1}^{(i)}}}{\vartheta(q_1;q_1)_{n_{k}^{(i)}-n_{k-1}^{(i)}}}\nonumber\\
    &\times \prod_{1\leq i \neq j \leq N} \prod_{k =1}^{N-i} \frac{\vartheta\big(q_1^{n_{k}^{(i)} - n_{k-j+i+1}^{(j)}}\Qs \frac{Q_{a_j}}{Q_{a_i}};q_1\big)_{n_{k}^{(i)}-n_{k-1}^{(i)}}\vartheta\big(q_1^{n_{k}^{(i)} - n_{N-k-j+i-1}^{(j)}-1} \Qs\frac{Q_{a_j}}{Q_{a_i}};q_1\big)_{n_{k}^{(i)}-n_{k-1}^{(i)}}}{\vartheta\big(q_1^{n_{k}^{(i)} - n_{k-j+i+1}^{(j)}} \frac{Q_{a_j}}{Q_{a_i}};q_1\big)_{n_{k}^{(i)}-n_{k-1}^{(i)}}\vartheta\big(q_1^{n_{k}^{(i)} - n_{N-k-j+i-1}^{(j)}-1} \frac{Q_{a_j}}{Q_{a_i}};q_1\big)_{n_{k}^{(i)}-n_{k-1}^{(i)}}}\,,
\end{align}
where we used the following parametrisation for $\lambda^{(1)},\ldots, \lambda^{(N)}$:
\begin{align}
    \lambda^{(i)}=(n_{N-i}^{(i)},\ldots,n_{1}^{(i)})\,,&& \forall i \in \{1,\ldots,N-1\}\,, && \lambda^{(N)}=\emptyset\,.
\end{align}
Our result matches the elliptic lift of the Noumi-Shiraishi representation of the MacDonald functions~\cite{Awata:2020yxf} up to a choice of parametrisation. After Higgsing of the left-over Coulomb branch moduli: $Q_{a_j} \to q_1^{\gamma_j}\Qs$, with $\gamma_j \in \mathbb{N}$ with $\gamma_j=2$, the expression~\eqref{eq:bulkdecouplinggeneralN} agrees with the result obtained in \cite{Koroteev:2019gqi,Gorsky:2021wio} and the elliptic analog of~\cite{Kimura:2022zsx}.

\section{Defect partition function of $\widehat{A}_{M-1}$ LSTs}\label{sec:genM}

\subsection{Double orbifold and character computation}

The setup discussed in section~\ref{sec:M=1} admits an interesting generalisation by considering an extra $\mathbb{Z}_M$ orbifold on the transverse space to the M$5$-branes. For the standard LST, this corresponds to the quiver generalisation given by~\figref{fig:Amquiver}. The M-theoretic brane construction is given by Table~\ref{tab:ZnZm}. 

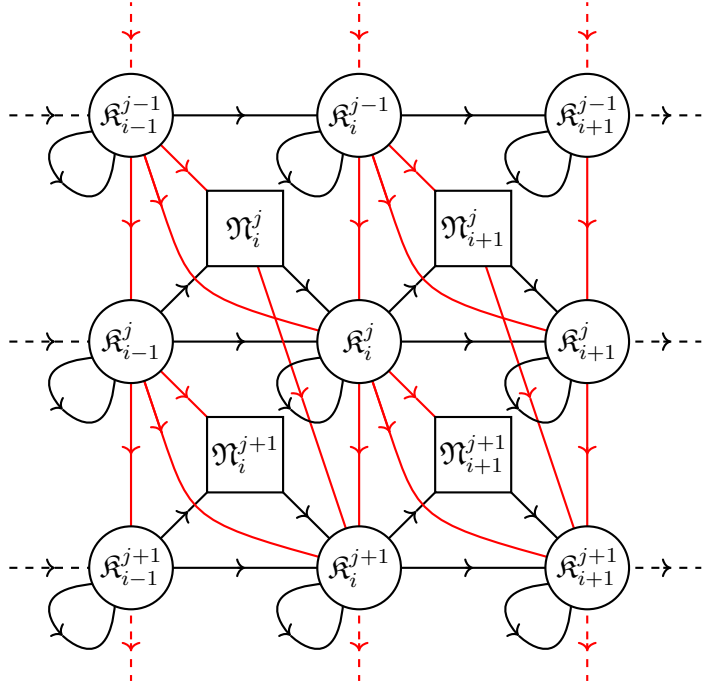
\begin{figure}[htbp]
    \centering
    \begin{tikzpicture}
        \draw[ thick,->] (-3,0) to (-1.5,0);
        \draw[ thick,->] (-1.6,0) to (0,0);
        \draw[ thick,->] (0,0) to (1.5,0);
        \draw[ thick,->] (1.4,0) to (3,0);
        \draw[ thick,->,color=red] (0,0) to (0,-1.5);
        \draw[ thick,->,color=red] (0,-1.4) to (0,-3);
        \draw[ thick,->,color=red] (3,0) to (3,-1.5);
        \draw[ thick,->,color=red] (3,-1.4) to (3,-3);
        \draw[ thick,->,color=red] (-3,0) to (-3,-1.5);
        \draw[ thick,->,color=red] (-3,-1.4) to (-3,-3);
        \draw[ thick,->,color=red] (0,3) to (0,1.5);
        \draw[ thick,->,color=red] (0,1.6) to (0,0);
        \draw[ thick,->,color=red] (-3,3) to (-3,1.5);
        \draw[ thick,->,color=red] (-3,1.6) to (-3,0);
        \draw[ thick,->,color=red] (3,3) to (3,1.5);
        \draw[ thick,->,color=red] (3,1.6) to (3,0);
        \draw[ thick,->] (-3,3) to (-1.5,3);
        \draw[ thick,->] (-1.6,3) to (0,3);
        \draw[ thick,->] (0,3) to (1.5,3);
        \draw[ thick,->] (1.4,3) to (3,3);
        \draw[ thick,->] (0,-3) to (1.5,-3);
        \draw[ thick,->] (1.4,-3) to (3,-3);
        \draw[ thick,->] (-3,-3) to (-1.5,-3);
        \draw[ thick,->] (-1.6,-3) to (0,-3);
        \draw[dashed, thick,->] (-4.6,-3) -- (-4,-3);
        \draw[dashed, thick,->] (-4,-3) -- (-3,-3);
        \draw[dashed, thick,->] (-4.6,0) -- (-4,0);
        \draw[dashed, thick,->] (-4,0) -- (-3,0);
        \draw[dashed, thick,->] (-4.6,3) -- (-4,3);
        \draw[dashed, thick,->] (-4,3) -- (-3,3);
        \draw[dashed, thick,->,color=red] (-3,-3) -- (-3,-4.1);
        \draw[dashed, thick,color=red] (-3,-4) -- (-3,-4.5);
        \draw[dashed, thick,->,color=red] (0,-3) -- (0,-4.1);
        \draw[dashed, thick,color=red] (0,-4) -- (0,-4.5);
        \draw[dashed, thick,->,color=red] (3,-3) -- (3,-4.1);
        \draw[dashed, thick,color=red] (3,-4) -- (3,-4.5);
        \draw[dashed, thick,color=red] (3,3) -- (3,4);
        \draw[dashed, thick,<-,color=red] (3,4) -- (3,4.5);
        \draw[dashed, thick,color=red] (0,3) -- (0,4);
        \draw[dashed, thick,<-,color=red] (0,4) -- (0,4.5);
        \draw[dashed, thick,color=red] (-3,3) -- (-3,4);
        \draw[dashed, thick,<-,color=red] (-3,4) -- (-3,4.5);
        \draw[dashed, thick,->] (3,3) -- (4.1,3);
        \draw[dashed, thick] (4,3) -- (4.5,3);
        \draw[dashed, thick,->] (3,0) -- (4.1,0);
        \draw[dashed, thick] (4,0) -- (4.5,0);
        \draw[dashed, thick,->] (3,-3) -- (4.1,-3);
        \draw[dashed, thick] (4,-3) -- (4.5,-3);
        \draw[thick,->] (-3,0) -- (-2.25,0.75);
        \draw[thick] (-2.25,0.75) -- (-1.5,1.5);
        \draw[thick,->] (-3,-3) -- (-2.25,-2.25);
        \draw[thick] (-2.25,-2.25) -- (-1.5,-1.5);
        \draw[thick,->] (0,0) -- (0.75,0.75);
        \draw[thick] (0.75,0.75) -- (1.5,1.5);
        \draw[thick,->] (0,-3) -- (0.75,-2.25);
        \draw[thick] (0.75,-2.25) -- (1.5,-1.5);
        \draw[thick,>-] (-0.75,0.75) -- (0,0);
        \draw[thick] (-1.5,1.5) -- (-0.7,0.7);
        \draw[thick,>-] (-0.75,-2.25) -- (0,-3);
        \draw[thick] (-1.5,-1.5) -- (-0.7,-2.3);
        \draw[thick,>-] (2.25,0.75) -- (3,0);
        \draw[thick] (1.5,1.5) -- (2.3,0.7);
        \draw[thick,>-] (2.25,-2.25) -- (3,-3);
        \draw[thick] (1.5,-1.5) -- (2.3,-2.3);
        \draw[thick,->,color=red] (-3,3) -- (-2.25,2.25);
        \draw[thick,color=red] (-2.25,2.25) -- (-1.5,1.5);
        \draw[thick,->,color=red] (-3,0) -- (-2.25,-0.75);
        \draw[thick,color=red] (-2.25,-0.75) -- (-1.5,-1.5);
        \draw[thick,->,color=red] (0,3) -- (0.75,2.25);
        \draw[thick,color=red] (0.75,2.25) -- (1.5,1.5);
        \draw[thick,->,color=red] (0,0) -- (0.75,-0.75);
        \draw[thick,color=red] (0.75,-0.75) -- (1.5,-1.5);
        \draw[thick,->,color=red] (-1.5,1.5) -- (-0.75,-0.75);
        \draw[thick,color=red] (-0.75,-0.75) -- (0,-3);
        \draw[thick,->,color=red] (1.5,1.5) -- (2.25,-0.75);
        \draw[thick,color=red] (2.25,-0.75) -- (3,-3);
        \draw[thick,->] (-0.2,-0.2) .. controls (-1.6,-0.2) and (-0.9,-0.9) .. (-0.9,-0.9 );
        \draw[thick] (-0.9,-0.9) .. controls (-0.9,-0.9) and (-0.2,-1.6) .. (-0.2,-0.2 );
        \draw[thick,->] (-3.2,-0.2) .. controls (-4.6,-0.2) and (-3.9,-0.9) .. (-3.9,-0.9 );
        \draw[thick] (-3.9,-0.9) .. controls (-3.9,-0.9) and (-3.2,-1.6) .. (-3.2,-0.2 );
        \draw[thick,->] (2.8,-0.2) .. controls (1.4,-0.2) and (2.1,-0.9) .. (2.1,-0.9 );
        \draw[thick] (2.1,-0.9) .. controls (2.1,-0.9) and (2.8,-1.6) .. (2.8,-0.2 );
        \draw[thick,->] (-0.2,2.8) .. controls (-1.6,2.8) and (-0.9,2.1) .. (-0.9,2.1);
        \draw[thick] (-0.9,2.1) .. controls (-0.9,2.1) and (-0.2,1.4) .. (-0.2,2.8);
        \draw[thick,->] (-3.2,2.8) .. controls (-4.6,2.8) and (-3.9,2.1) .. (-3.9,2.1);
        \draw[thick] (-3.9,2.1) .. controls (-3.9,2.1) and (-3.2,1.4) .. (-3.2,2.8 );
        \draw[thick,->] (2.8,2.8) .. controls (1.4,2.8) and (2.1,2.1) .. (2.1,2.1 );
        \draw[thick] (2.1,2.1) .. controls (2.1,2.1) and (2.8,1.4) .. (2.8,2.8 );
        \draw[thick,->] (-0.2,-3.2) .. controls (-1.6,-3.2) and (-0.9,-3.9) .. (-0.9,-3.9 );
        \draw[thick] (-0.9,-3.9) .. controls (-0.9,-3.9) and (-0.2,-4.6) .. (-0.2,-3.2 );
        \draw[thick,->] (-3.2,-3.2) .. controls (-4.6,-3.2) and (-3.9,-3.9) .. (-3.9,-3.9 );
        \draw[thick] (-3.9,-3.9) .. controls (-3.9,-3.9) and (-3.2,-4.6) .. (-3.2,-3.2 );
        \draw[thick,->] (2.8,-3.2) .. controls (1.4,-3.2) and (2.1,-3.9) .. (2.1,-3.9 );
        \draw[thick] (2.1,-3.9) .. controls (2.1,-3.9) and (2.8,-4.6) .. (2.8,-3.2 );
        \draw[thick,->,color=red] (0,0) .. controls (0.8,-2.4) and (0,0) .. (0.4,-1.2 );
        \draw[thick,->,color=red] (0,0) .. controls (0.8,-2.4) and (0,0) .. (0.4,-1.2 ) ..controls (0.8,-2.4) and (0.8,-2.4) .. (3,-3);
        \draw[thick,->,color=red] (-3,0) .. controls (-2.2,-2.4) and (-3,0) .. (-2.6,-1.2 );
        \draw[thick,->,color=red] (-3,0) .. controls (-2.2,-2.4) and (-3,0) .. (-2.6,-1.2 ) ..controls (-2.2,-2.4) and (-2.2,-2.4) .. (0,-3);
        \draw[thick,->,color=red] (0,3) .. controls (0.8,0.6) and (0,3) .. (0.4,1.8 );
        \draw[thick,->,color=red] (0,3) .. controls (0.8,0.6) and (0,3) .. (0.4,1.8) ..controls (0.8,0.6) and (0.8,0.6) .. (3,0);
        \draw[thick,->,color=red] (-3,3) .. controls (-2.2,0.6) and (-3,3) .. (-2.6,1.8);
        \draw[thick,->,color=red] (-3,3) .. controls (-2.2,0.6) and (-3,3) .. (-2.6,1.8) ..controls (-2.2,0.6) and (-2.2,0.6) .. (0,0);
        \quivbox{1.5}{1.5}{0.5}
        \quivbox{-1.5}{1.5}{0.5}
        \quivbox{-1.5}{-1.5}{0.5}
        \quivbox{1.5}{-1.5}{0.5}
        \draw[ thick,fill=white] (0,0) circle (0.55cm);
        \draw[ thick,fill=white] (3,0) circle (0.55cm);
        \draw[ thick,fill=white] (-3,0) circle (0.55cm);
        \draw[ thick,fill=white] (0,3) circle (0.55cm);
        \draw[ thick,fill=white] (0,-3) circle (0.55cm);
        \draw[ thick,fill=white] (3,3) circle (0.55cm);
        \draw[ thick,fill=white] (-3,-3) circle (0.55cm);
        \draw[ thick,fill=white] (3,-3) circle (0.55cm);
        \draw[ thick,fill=white] (-3,3) circle (0.55cm);
        \node at (0,0) {$\mathfrak{K}_i^j$};
        \node at (0,3) {$\mathfrak{K}_{i}^{j-1}$};
        \node at (0,-3) {$\mathfrak{K}_{i}^{j+1}$};
        \node at (3,0) {$\mathfrak{K}_{i+1}^{j}$};
        \node at (3,3) {$\mathfrak{K}_{i+1}^{j-1}$};
        \node at (3,-3) {$\mathfrak{K}_{i+1}^{j+1}$};
        \node at (-3,-3) {$\mathfrak{K}_{i-1}^{j+1}$};
        \node at (-3,0) {$\mathfrak{K}_{i-1}^{j}$};
        \node at (-3,3) {$\mathfrak{K}_{i-1}^{j-1}$};
        \node at (1.5,1.5) {$\mathfrak{N}_{i+1}^j$};
        \node at (-1.5,1.5) {$\mathfrak{N}_{i}^j$};
        \node at (-1.5,-1.5) {$\mathfrak{N}_{i}^{j+1}$};
        \node at (1.5,-1.5) {$\mathfrak{N}_{i+1}^{j+1}$};
    \end{tikzpicture}
    \caption{A portion of the doubly cyclic chain-saw ADHM quiver associated with the $\mathbb{Z}_N \times \mathbb{Z}_M$-orbifolded instanton moduli space. Red arrows relate different sector of the $\mathbb{Z}_M$ orbifold and black arrows relate different sectors of the $\mathbb{Z}_N$ orbifold.}
    \label{fig:orrbichainsaw}
\end{figure}

\begin{table}[htbp]
    \centering
    \begin{tabular}{c||c|c|c|c|c|c}
         & $\mathbb{C}_1$ &$\mathbb{C}_2/\mathbb{Z}_N$ & $\mathbb{T}^2$ & $\mathbb{S}^1_{\perp}$ & $\mathbb{C}_3/\mathbb{Z}_M$ & $\mathbb{C}_4/(\mathbb{Z}_N\times \mathbb{Z}_M)$  \\
         \hline\hline
       $N$ M$5$  & -\,- & -\,- & -\,- & $\times$ & &\\
       $k$ M$2$ &  & & -\,- & - & & \\
     \end{tabular}
    \caption{M$5$/M$2$ setup engineering a $\widehat{A}_{M-1}$ quiver LST with rank $N$ node in presence of a codimension-2 surface defect.}
    \label{tab:ZnZm}
\end{table}
We directly consier the case of the full type defect, the complete orbifold action on the spacetime coordinate $(z_1,z_2,z_3,z_4)\in \mathbb{C}_1 \times \mathbb{C}_2 \times \mathbb{C}_3 \times \mathbb{C}_4$ is:
\begin{align}
    \mathbb{Z}_N \times \mathbb{Z}_M : \begin{pmatrix}
        z_1\\z_2\\z_3\\ z_4
    \end{pmatrix} \longrightarrow \begin{pmatrix}
        1 & 0 &0 &0 \\
        0 & e^{\frac{2i\pi}{N}} & 0 &0\\
        0 & 0 & e^{\frac{2i\pi}{M}} & 0\\
        0 & 0 & 0 & e^{-\frac{2i\pi}{N}-\frac{2i\pi}{M}}
    \end{pmatrix}\cdot \begin{pmatrix}
        z_1\\z_2\\z_3\\ z_4
    \end{pmatrix}\,.
\end{align}
\paragraph{}
Using this setup, we can naturally generalise the character computation of subsection~\ref{subsec:characcomM1} by using the formalism of double quiver gauge theories~\cite{Kimura:2022zsm}. Similarly, we can characterise the moduli space of instanton associated with this configuration by decomposing the framing and instanton vector spaces on irreducible representations of $\mathbb{Z}_N$ and $\mathbb{Z}_{M}$:
\begin{align}
    &\mathfrak{N} = \bigoplus_{i=1}^N \bigoplus_{j=1}^M \mathfrak{N}_{i}^j \otimes \mathfrak{R}_{i,N} \otimes \mathfrak{R}_{j,M}\,, && {\rm dim}_{\mathbb{C}}\, \mathfrak{ N}_{i}^j = N\,,\nonumber\\
    &\mathfrak{K} = \bigoplus_{i=1}^N \bigoplus_{j=1}^M \mathfrak{K}_{i}^j \otimes \mathfrak{R}_{i,N} \otimes \mathfrak{R}_{j,M}\,, && {\rm dim}_{\mathbb{C}}\, \mathfrak{ K}_{i}^j = k_i^j\,.
\end{align}
where $\{\mathfrak{R}_{i,P}\}_{i\in \{1,\ldots,P\}}$ are the irreducible representations of $\mathbb{Z}_P$. The cotangent bundle to the origin of the space-time decomposes as:
\begin{align}
    \mathfrak{Q} = \bigoplus_{i=1}^4 \mathfrak{Q}_i = T_o^\vee (\mathbb{C}_1 \times \mathbb{C}_2/\mathbb{Z}_N \times \mathbb{C}_3/\mathbb{Z}_M \times \mathbb{C}_4/(\mathbb{Z}_N \times \mathbb{Z}_M))\,.
\end{align}
We define the character of the framing and instanton vector spaces in the following manner:
\begin{align}
    &{\bf N}_i^j:={\rm ch}\, \mathfrak{ N}_{i}^j = \sum_{\alpha=1}^NQ_{a_{i,\alpha}^j} \,, && \forall i \in \{1,\ldots,N\}\,, && \forall j \in \{1, \ldots,M\}\,,\nonumber\\
    &{\bf K}_i^j:={\rm ch}\, \mathfrak{K}_{i}^j = \sum_{I=1}^{k_i^j}Q_{\phi_{i,I}^j} \,, && \forall i \in \{1,\ldots,N\}\,, && \forall j \in \{1, \ldots,M\}\,,
\end{align}
and the matrix-valued characters of the $\mathfrak{Q}_i$ for $i\in\{1,\ldots,4\}$ as:
\begin{align}
    &{\bf Q}_1:={\rm ch}\, \mathfrak{Q}_1 = q_1 \mathds{1}_{N} \otimes \mathds{1}_{M} \,, \hspace{4cm} q_1:=e^{2i\pi \varepsilon_1}\,, \nonumber\\
    &{\bf Q}_2:={\rm ch}\, \mathfrak{Q}_2 = \tq \mathcal{R}_{1,N}\otimes \mathds{1}_{M}\,, \hspace{4cm} \tq := e^{\frac{2i\pi \varepsilon_2}{N}}\,,\nonumber\\
    &{\bf Q}_3:={\rm ch}\, \mathfrak{Q}_3 = Q_{\widehat{S}}^{-1} \mathds{1}_N \otimes \mathcal{R}_{1,M}\,,  \hspace{4cm} Q_{\widehat{S}} := e^{\frac{2i\pi S}{M}}\,, \nonumber\\
    &{\bf Q}_4:= {\rm ch}\, \mathfrak{Q}_4 = Q_{\widehat{S}} q_1^{-1}\tq^{-1} \mathcal{R}_{N-1,N} \otimes \mathcal{R}_{M-1,M}\,.
\end{align}
satisfying ${\bf Q}_1{\bf Q}_2{\bf Q}_3{\bf Q}_4 = \mathds{1}_N\otimes \mathds{1}_M $ where $\{\mathcal{R}_{i,P}\}_{i\in\{1 \ldots P\}}$ are matrix representations of $\{\mathfrak{R}_{i,P}\}_{i\in\{1 \ldots P\}}$ which are given by:
\begin{align}
    [\mathcal{R}_{1,P}]_{kl} = \delta_{k+1,l}\,, && \forall s \in \mathbb{Z}\,, &&\mathcal{R}_{s,P} = \mathcal{R}_{1,P}^s\,, && \mathcal{R}_{0,P}=\mathcal{R}_{P,P}=\mathds{1}_P\,,
\end{align}
and we define the following convenient notations:
\begin{align}
    {\bf P}_i := \mathds{1}_{N}\otimes \mathds{1}_M - {\bf Q}_i\,, && {\bf P}_{ij} = {\bf P}_i {\bf P}_j\,.
\end{align}
We then follow the same derivation as in section~\ref{subsec:combexprM=1} of the character of the instanton moduli space and define the character of the observable sheaf and the vector multiplet analogue as:
\begin{align}
    {\bf Y}:= {\bf P}_{34} [{\bf N}-{\bf P}_{12} {\bf K}]\,, && {\bf V}:={\bf Y}^\vee {\bf P}_{12}^{-1} {\bf P}_{34}^{-1} {\bf Y}={\bf v} + {\bf v}^\vee\,,
\end{align}
we can then extract the relevant characters from $\mathbb{Z}_N\times \mathbb{Z}_M$-invariant sub-sector of $\bf v$ defined as:
\begin{align}\label{eq:chiM}
    &\chi =: \chi_{\rm pert.} + \chi_{\rm inst.}\,, \hspace{4cm}  \chi_{\rm pert.}= \big[{\bf P}_3^\vee {\bf N}^\vee {\bf P}_{12}^{-1} {\bf N} \big]_{\mathbb{Z}_N \times \mathbb{Z}_M}\,,\nonumber\\
    & \chi_{\rm inst.}=\big[{\bf P}_3(- {\bf N}^\vee {\bf K} - {\bf Q}_1^\vee {\bf Q}_2^\vee {\bf K}^\vee {\bf N}+ {\bf P}_{12}^\vee {\bf K}^\vee {\bf K} ) \big]_{\mathbb{Z}_N \times \mathbb{Z}_M}\,,
\end{align}
with the $\mathbb{Z}_N\times \mathbb{Z}_M$-invariant sub-sector defined as the trace on the matrix-valued characters:
\begin{align}
    \left[ \sum_{i=1}^N \sum_{j=1}^M c_i^j\, \mathcal{R}_{i,N} \otimes \mathcal{R}_{j,M} \right]_{\mathbb{Z}_N \times \mathbb{Z}_M} = \frac{1}{NM} {\rm Tr}\left[ \sum_{i=1}^N \sum_{j=1}^M c_i^j\, \mathcal{R}_{i,N} \otimes \mathcal{R}_{j,M} \right] \,, && \forall c_i^j \in \mathbb{C}\,.
\end{align}
The invariant sub-sector of the character of the tangent space to the instanton moduli space is therefore given by:
\begin{align}\label{eq:chiinst}
    &\chi_{\rm inst.} = \sum_{j=1}^M \sum_{i=1}^N \big[ - ({\bf N}_i^{j})^\vee {\bf K}_i^j - q_1^{-1} \tq^{-1} ({\bf K}_{i-1}^{j})^\vee {\bf N}_i^j + (1-q_1^{-1}) ({\bf K}_i^{j})^\vee {\bf K}_i^j - \tq^{-1} (1-q_1^{-1}) ({\bf K}_{i-1}^{j})^\vee {\bf K}_i^j \nonumber\\
    &+ Q_{\widehat{S}}\big(({\bf N}_i^{j-1})^\vee {\bf K}_i^j + q_1^{-1} \tq^{-1}  ({\bf K}_{i-1}^{j-1})^\vee {\bf N}_i^j - (1-q_1^{-1})({\bf K}_i^{j-1})^\vee {\bf K}_i^j + \tq^{-1} (1-q_1^{-1}) ({\bf K}_{i-1}^{j-1})^\vee {\bf K}_i^j \big) \big]\,.
\end{align}
This result corresponds to the ADHM equations of a new doubly periodic quiver given by~\figref{fig:orrbichainsaw} which corresponds to a generalisation of the chain-saw quiver of~\cite{Kanno:2011fw}. 

\subsection{Combinatoric expression}

Taking the $6$d index~\eqref{eq:6dindex} of the character of the tangent space to the instanton moduli space leads to the LMNS formula:
\begin{align}
    Z_{\rm def.}^{\underline{\underline{k}},N,M} = \frac{1}{\underline{\underline{k}}!} \frac{1}{[-\varepsilon_1]^{|\underline{\underline{k}}|}} \oint &\prod_{i=1}^N \prod_{j=1}^M \prod_{I=1}^{k_i^j} {\rm d}\phi_{i,I}^j \prod_{1\leq I < J \leq k_i^j} \frac{[\phi_{i,I}^j - \phi_{i,J}^j]}{[\phi_{i,I}^j - \phi_{i,J}^j-\varepsilon_1]} \prod_{I=1}^{k_{i-1}^j}\prod_{J=1}^{k_{i}^j} \frac{[\phi_{i,J}^j - \phi_{i-1,I}^j - \varepsilon_{1}-\frac{\varepsilon_2}{N}]}{[\phi_{i,J}^j - \phi_{i-1,I}^j - \frac{\varepsilon_{2}}{N}]} \nonumber\\
    &\times\prod_{j=1}^M \prod_{i=1}^N \prod_{\alpha=1}^N \frac{\prod_{I=1}^{k_i^{j-1}}[\widehat{S}+\phi_{i,I}^{j-1} - a_\alpha^{j-1}]\prod_{I=1}^{k_{i-1}^{j-1}}[\widehat{S}+a_{i,\alpha}^{j-1}-\phi_{i-1,I}^{j-1}-\varepsilon_{1}-\frac{\varepsilon_2}{N}]}{\prod_{I=1}^{k_i^j}[\phi_{i,I}^j - a_{i,\alpha}^j]\prod_{I=1}^{k_{i-1}^j}[a_\alpha^j-\phi_{i-1,I}^j-\varepsilon_{1}-\frac{\varepsilon_2}{N}]}\nonumber\\
    & \times \prod_{I=1}^{k_i^{j-1}} \prod_{J=1}^{k_i^j} \frac{[\widehat{S} + \phi_{i,J}^{j-1} - \phi_{i,I}^j - \varepsilon_1]}{[\widehat{S} + \phi_{i,J}^{j-1} - \phi_{i,I}^j]} \prod_{I=1}^{k_{i-1}^{j-1}} \prod_{J=1}^{k_i^j} \frac{[\widehat{S} + \phi_{i-1,J}^{j-1} - \phi_{i,I}^j - \varepsilon_1]}{[\widehat{S} + \phi_{i-1,J}^{j-1} - \phi_{i,I}^j - \varepsilon_{1}-\frac{\varepsilon_2}{N}]}\,.
\end{align}
where $\underline{\underline{k}}:=\{k_i^j\}_{i\in\{1,\ldots,N\},j\in\{1,\ldots,M\}}$ and we use the same notations as in subsection~\ref{subsec:combexprM=1}. We choose an ordering of the integration in term of the charges $\underline{\underline{k}}$ which preserves the two cyclic symmetries:
\begin{align}
    \oint \prod_{i=1}^N \prod_{j=1}^M \prod_{I=1}^{k_i^j} \mathrm{d}\phi_{i,I}^j \longrightarrow \prod_{i=1}^N \prod_{j=1}^M \oint \mathrm{d} \phi_{i,\max(\underline{\underline{k}})}^j \ldots \prod_{i=1}^N \prod_{j=1}^M \oint \mathrm{d} \phi_{i,1}^j\,,
\end{align}
the pole structure is then governed $\forall i \in \{1,\ldots,N\}$ and $\forall j \in \{1,\ldots,M\}$ by the three relations:
\begin{align}\label{eq:polesAm}
    \phi_{i,I}^j = a_{i,\alpha}^j\,, && \phi_{i,J}^j = \phi_{i,I}^j + \varepsilon_1\,,\quad I>J\geq1\,, && \phi_{i,J}^j = \phi_{i-1,I}^j + \frac{\varepsilon_2}{N}\,,\quad I,J\geq1\,.
\end{align}
We observe that the different sectors under the $\mathbb{Z}_M$ orbifold do not mix, therefore configurations are classified by a $NM$-tuple of partitions $\underline{\underline{\lambda}}:= \{\lambda^{(i,j)}\}_{i\in\{1,\ldots,N\},j\in\{1,\ldots,M\}}$ such that:
\begin{align}
    {\bf K}_{i,k}^j \big|_{\underline{\underline{\lambda}}} := \begin{cases}
    \tq^{-1} Q_{a_{i-k+1}^j} \omega_{k|N}( \sum_{(n,m)\in \lambda^{(i-k+1,j)}} q_1^{m-1} \tq^n) \quad \quad \text{if} \quad k \leq i\,,\\
     Q_{a_{i-k+1}^j} \omega_{k|N}( \sum_{(n,m)\in \lambda^{(i-k+1,j)}} q_1^{m-1} \tq^n) \quad \quad \text{other}\,,
\end{cases}
\end{align}
which is related to the ${\bf K}_i^j\big|_{\underline{\underline{\lambda}}}$ by:
\begin{align}
    {\bf K}_i^j\big|_{\underline{\underline{\lambda}}} = \sum_{k=1}^N {\bf K}_{i,k}^j\big|_{\underline{\underline{\lambda}}}\,.
\end{align}
The colouring function is independent of the $\mathbb{Z}_M$ orbifold sector and is therefore related to~\eqref{eq:colorf} by
\begin{align}
k_i^j(\underline{\underline{\lambda}})=k_i(\underline{\lambda}^{(j)})\,, \quad \quad \underline{\lambda}^{(j)} :=\left\{\lambda^{(1,j)},\ldots,\lambda^{(N,j)}\right\}\,.
\end{align}
We can then use these expressions and the properties of $\omega_{p|N}$ in Appendix~\ref{subsec:propome} to obtain a combinatoric expression for the defect PF, we refer the reader to subsection~\ref{subsec:combexprM=1} for a more detailed discussion of this derivation. Using the notation $\underline{\underline{X}}:=\{X_i^j\}_{i\in\{1\ldots N\},j\in\{1\ldots M\}}$, we obtain:
\begin{tcolorbox}[colback=black!10!white,colframe=black!95!green]
\begin{align}
    \mathcal{Z}_{\rm def.}^{\rm inst.,(N,M)}(\underline{\underline{\mathfrak{q}}},\mathfrak{q}|\underline{\underline{Q_a}},\Qr|Q_{\widehat{S}},q_1,\tq)=&\sum_{\underline{\underline{\lambda}}} \left[\prod_{i=1}^N \prod_{j=1}^M \left(\mathfrak{q}_{i,j} \right)^{k_i^j(\underline{\underline{\lambda}})} \right] \nonumber\\
    &\times\prod_{l=1}^M \prod_{1\leq i,j\leq N} \frac{\mathcal{N}_{\lambda^{(i,l-1)}\lambda^{(j,l)}}^{(j-i|N)}(Q_{\widehat{S}} Q_{a_j^l}/Q_{a_i^{l-1}},\Qr;q_{1},\tq)}{\mathcal{N}_{\lambda^{(i,l)}\lambda^{(j,l)}}^{(j-i|N)}(Q_{a_j^l}/Q_{a_i^{l}},\Qr;q_{1},\tq)}\,\label{ZinstM}
\end{align}
\end{tcolorbox}
\noindent
where the orbifolded Nekrasov subfunctions $\mathcal{N}^{(k|N)}$ are defined by~\eqref{eq:Np} and $\mathfrak{q}_{i,j}$ is the instanton counting parameter associated with the $\mathfrak{K}_i^j$ vector space. Alternatively, it can be interpreted as the exponentiated gauge coupling of the $i$-th $U(1)$ gauge group from the breaking of the $j$-th $U(N)$ in the cyclic quiver. As such, it satisfies two relations:
\begin{align}\label{eq:paramtauM}
    \mathfrak{q}_j = \prod_{i=1}^N \mathfrak{q}_{i,j}\,, && \mathfrak{q}= \prod_{j=1}^N \mathfrak{q}_{j}\,, && \mathfrak{q}=e^{2i\pi \tau}\,, && \mathfrak{q}_j=e^{2i\pi \tau_j}&& \mathfrak{q}_{i,j}=e^{2i\pi\tau_{i,j}}\,,
\end{align}
where $\tau_j$ is the gauge coupling of the $j$-th $U(N)$ gauge node in the cyclic quiver and $\tau$ can be thought as the affinisation parameter of the $\widehat{A}_{M-1}$ Dynkin diagram and directly related to the radius of one of the two compact space-time dimensions.

\paragraph{}
For later convenience, we introduce the shorthand notation:
\begin{align}
    Z[\underline{\underline{\lambda}}] = Z \begin{bmatrix}
        \lambda^{(1,1)} & \cdots & \lambda^{(N,1)}\\
        \vdots &\ddots & \vdots \\
        \lambda^{(1,M)} & \cdots & \lambda^{(N,M)}
    \end{bmatrix}:= \prod_{l=1}^M \prod_{1\leq i,j\leq N} \frac{\mathcal{N}_{\lambda^{(i,l-1)}\lambda^{(j,l)}}^{(j-i|N)}(Q_{\widehat{S}} Q_{a_j^l}/Q_{a_i^{l-1}},\Qr;q_{1},\tq)}{\mathcal{N}_{\lambda^{(i,l)}\lambda^{(j,l)}}^{(j-i|N)}(Q_{a_j^l}/Q_{a_i^{l}},\Qr;q_{1},\tq)}\,,
\end{align}
with $\underline{\underline{\lambda}}=\{\lambda^{(i,j)}\}_{i\in\{1,\ldots,N\},j\in \{1,\ldots,M\}}$.

\subsection{Perturbative contribution}
We now briefly detail the computation of the perturbative contribution to the full defect PF. From~\eqref{eq:chiM}, the perturbative character is given by:
\begin{align}
    \chi_{\rm pert.} = \sum_{j=1}^M \sum_{k=1}^N \sum_{l=1}^N ({\bf N}_{k}^j)^\vee \frac{1}{1-q_1} \frac{\tq^l}{1-\tq^N} {\bf N}_{k+l}^{j} - Q_{\widehat{S}}\sum_{j=1}^M \sum_{k=1}^N \sum_{l=1}^N ({\bf N}_{k}^{j-1})^\vee \frac{1}{1-q_1} \frac{\tq^l}{1-\tq^N} {\bf N}_{k+l}^{j}\,,
\end{align}
where similarly to the $M=1$ case, we used~\eqref{eq:expansionP2}. Taking the $6$d index~\eqref{eq:6dindex} of this character, we obtain the perturbative contribution to the defect PF:
\begin{align}
    \mathcal{Z}_{\rm def.}^{\rm pert.,(N,M)}(\underline{\underline{Q_a}},\Qr;\Qs,q_1,\tq):= \prod_{j=1}^M \prod_{1\leq k < l \leq N} &\frac{\vartheta\bigg(\frac{Q_{a_l^j}}{Q_{a_k^{j-1}}}Q_{\widehat{S}} \tq^{l-k};q_1,\tq^N\bigg)_\infty}{\vartheta\bigg(\frac{Q_{a_l^j}}{Q_{a_k^{j}}}\tq^{l-k};q_1,\tq^N\bigg)_\infty}\nonumber\\
    &\times \prod_{1\leq l \leq k \leq N} \frac{\vartheta\bigg(\frac{Q_{a_k^j}}{Q_{a_l^{j-1}}}Q_{\widehat{S}} \tq^{N-l+k};q_1,\tq^N\bigg)_\infty}{\vartheta\bigg(\frac{Q_{a_k^j}}{Q_{a_l^{j}}}\tq^{N-l+k};q_1,\tq^N\bigg)_\infty}\,,
\end{align}
with the Pochhammer symbol defined by~\eqref{eq:appellpochdef}. The full defect PF is then given:
\begin{align}
    \mathcal{Z}_{\rm def.}^{\rm (N,M)}(\underline{\underline{\frak q}},\mathfrak{q};\underline{\underline{Q_a}},\Qr;\Qs,q_1,\tq):=\mathcal{Z}_{\rm def.}^{\rm pert.,(N,M)}(\underline{\underline{Q_a}},\Qr;\Qs,q_1,\tq) \mathcal{Z}_{\rm def.}^{\rm inst.,(N,M)}(\underline{\underline{\frak q}},\mathfrak{q};\underline{\underline{Q_a}},\Qr;\Qs,q_1,\tq).
\end{align}

\subsection{Non-perturbative symmetries}\label{subsec:npsymNM}

\paragraph{}
The results on non-perturbative symmetries obtained for the case $M=1$ in subsection~\ref{subsec:npsymNM1} admit simple generalisations for any $M$. The vector parametrising the moduli space of the $A$-type LST with a surface defect is given by
\begin{align}
    \vec v_{\rm def.}^{(N,M)}=(\{a_i^{j}\}_{i\in\{1\ldots N-1 \},j\in\{1\ldots M\}},\{\tau_{i,1}\}_{i\in \{1\ldots N\}},\ldots,\{\tau_{i,M}\}_{i\in \{1\ldots N\}},S,\rho)\,,
\end{align}
which can be seen as the extension of the vector parametrising the moduli space of the $A$-type LST~\cite{Filoche:2023yfm}\footnote{We used a slightly different parametrisation than in \cite{Filoche:2023yfm} since we parametrise the gauge couplings by $\tau_{i\in \{1,\ldots,M\}}$ instead of $\tau_{i\in \{1,\ldots,M-1\}},\tau$, the relation between the two is simply given by~\eqref{eq:paramtauM}.}:
\begin{align}
    \vec v^{(N,M)} = (\{\widehat{a}_i^{(j)}\}_{i\in\{1\ldots N-1 \},j\in\{1\ldots M\}},\tau_1,\ldots,\tau_M,S,\rho)\,,
\end{align}
where the change of variables between $\vec v_a=\{a_i^{j}\}_{i\in\{1\ldots N-1 \},j\in\{1\ldots M\}}$ and $\vec v_{\widehat{a}}=\{\widehat{a}_i^{(j)}\}_{i\in\{1\ldots N-1 \},j\in\{1\ldots M\}}$ is given by the relations:
\begin{align}
    Q_{a_j^{l-1}}/Q_{a_i^l} = \widehat{Q}_{j,j-i}^{(l-1)} Q_{\widehat{S}}^{-1}\,, && Q_{a_j^l}/Q_{a_i^l}= \overline{Q}_{j,j-i}^{(l)}\,,&& \forall i,j,l, && 1\leq i,j \leq N\,, && 1\leq l \leq M\,,
\end{align}
where the functions $\widehat{Q}$ and $\overline{Q}$ are defined as functions of the $\{\widehat{a}_i^{(j)}\}_{i\in\{1\ldots N-1 \},j\in\{1\ldots M\}}$ parameters following (2.10) of \cite{Filoche:2023yfm}. Solving this relations, we define the change of basis by:
\begin{align}
    \vec v_a = \mathcal{P}_{a \widehat{a}} \cdot \vec v_{\widehat{a}}\,, && \mathcal{P}_{a \widehat{a}} \in \mathbb{M}_{(N-1)M \times (N-1)M}(\mathbb{Q})\,,
\end{align}
The set symmetries of the theory without defects classified in \cite{Filoche:2023yfm} $\mathbb{T}=\{\mathfrak{t}_i\}_{i \in \{1,\ldots,6\}}$ can be parametrised as:
\begin{align}
    &\mathfrak{t}_i : (\vec v^{(N,M)})^T \longrightarrow \mathcal{T}_i \cdot(\vec v^{(N,M)})^T\,, \quad \quad \quad  \quad \mathcal{T}_i = \begin{pmatrix}
        \mathcal{T}_{i,\widehat{a}\widehat{a}}& \mathcal{T}_{i,\widehat{a}\tau} & \mathcal{T}_{i,\widehat{a}\omega}\\
        \mathcal{T}_{i,\tau\widehat{a}}& \mathcal{T}_{i,\tau\tau} & \mathcal{T}_{i,\tau\omega}\\
        \mathcal{T}_{i,\omega\widehat{a}}& \mathcal{T}_{i,\omega\tau} & \mathcal{T}_{i,\omega\omega}
    \end{pmatrix}\,,\\
    &\mathcal{T}_{i,\widehat{a}\widehat{a}} \in \mathbb{M}_{(N-1)M\times(N-1)M}(\mathbb{Q})\,, \quad \quad \mathcal{T}_{i,\tau\tau} \in \mathbb{M}_{M\times M}(\mathbb{Q})\,, \quad \quad T_{i,\omega \omega}\in \mathbb{M}_{2\times2}(\mathbb{Q})\,,
\end{align}
which can be extended into symmetries of the theory with defect in the following way:
\begin{align}
    &\mathfrak{t}_{\rm def.,\mathit{i}} : (\vec v^{(N,M)}_{\rm def.})^T \longrightarrow \mathcal{T}_{\rm def.,\mathit{i}} \cdot(\vec v^{(N,M)}_{\rm def.})^T\,,\nonumber\\
    &\mathcal{T}_{\rm def.,\mathit{i}} = \begin{pmatrix}
        \mathcal{P}_{a \widehat{a}}\mathcal{T}_{i,\widehat{a}\widehat{a}}\mathcal{P}_{a \widehat{a}}^{-1}& \frac{1}{N} \mathcal{P}_{a \widehat{a}}\mathcal{T}_{i,\widehat{a}\tau}\otimes \mathbb{J}_{1,N} & \mathcal{P}_{a \widehat{a}}\mathcal{T}_{i,\widehat{a}\omega}\\
        \frac{1}{N}(\mathcal{T}_{i,\tau\widehat{a}}\otimes \mathbb{J}_{N,1}) \mathcal{P}_{a \widehat{a}}^{-1}& \frac{1}{N}\mathcal{T}_{i,\tau\tau}\otimes \mathbb{J}_{N,N} &\frac{1}{N} \mathcal{T}_{i,\tau\omega}\otimes \mathbb{J}_{N,1} \\
        \mathcal{T}_{i,\omega\widehat{a}}\mathcal{P}_{a \widehat{a}}^{-1}& \frac{1}{N} \mathcal{T}_{i,\omega\tau}\otimes \mathbb{J}_{1,N} & \mathcal{T}_{i,\omega\omega}
    \end{pmatrix}\,,
\end{align}
with the definition:
\begin{align}
    &\mathbb{J}_{p,q} = \begin{pmatrix}
        1 & \cdots & 1 \\ 
        \vdots & \ddots & \vdots \\
        1 & \cdots & 1
    \end{pmatrix} \in \mathbb{M}_{p,q}(\mathbb{N})\,.
\end{align}

\subsection{NS limit}

\paragraph{}
In this subsection, we discuss the regularity of the defect partition function in the NS limit, thereby extending the results obtained in the case $M=1$. We make the same observations as in subsection~\ref{subsec:NSgenN}, the defect partition function admits two natural normalisations:
\begin{itemize}
    \item The defect instanton PF can be regularized by the bulk instanton PF which can be obtained as a restriction of the defect PF:
    \begin{align}\label{eq:ZbulkM}
        \mathcal{Z}_{\rm bulk.}^{\rm inst.,(N,M)} := \sum_{\tiny \begin{matrix}\underline{\lambda}^{(j)} \in \big(\mathcal{P}_{\rm bulk}^{(N)} \big)^N \\ j\in \{1,\ldots,M \}\end{matrix} } \prod_{i=1}^N \prod_{j=1}^M \mathfrak{q}_{i,j}^{k_i^j(\underline{\underline{\lambda}})} Z[\underline{\underline{\lambda}}]\,,
    \end{align}
    with $\mathcal{P}_{\rm bulk}^{(N)}$ defined by~\eqref{eq:defPbulkN} and $\underline{\lambda}^{(j)}=\{\lambda^{(i,j)}\}_{i\in\{1,\ldots,N\}}$.
    \item The defect instanton PF can be regularized by a generalisation of Shiraishi's normalizer defined by:
    \begin{align}\label{eq:ZfrakM}
        \mathfrak{Z}^{(N,M)} := \sum_{\tiny \begin{matrix}
            \underline{\lambda}^{(j)} \in \mathcal{P}^N, \, j\in\{1,\ldots,M\}\\
            k_1(\underline{\lambda}^{(j)})=\cdots = k_N(\underline{\lambda}^{(j)})
        \end{matrix}} \prod_{i=1}^N \prod_{j=1}^M \mathfrak{q}_{i,j}^{k_i^j(\underline{\underline{\lambda}})} Z[\underline{\underline{\lambda}}]\,.
    \end{align}
    \item The defect PF can be regularized by any normaliser $\mathcal{N}^{(N,M)}$ which can be written in the following way:
    \begin{align}
        \mathcal{N}^{(N,M)} = \mathcal{Z}_{\rm bulk}^{\rm inst.,(N,M)} + \sum_{\tiny \begin{matrix}\underline{\mu}^{(j)} \in \mathcal{S}^{(j)}\subset  \big (\mathcal{\mathcal{P}}_{\rm seed}^{(N)} \big)^N\\ j \in \{1,\ldots,N\} \end{matrix}} \sum_{\tiny \begin{matrix}\underline{\nu}^{(j)} \in \mathcal{T}^{(N)}(\underline{\mu}^{(j)})\\j \in \{1,\ldots,N\} \end{matrix}} \prod_{i=1}^N \prod_{j=1}^M \mathfrak{q}_{i,j}^{k_i^j(\underline{\underline{\nu}})} Z[\underline{\underline{\nu}}]\,,
    \end{align}
    with $\mathcal{S}^{(j)}$ is any subset of $(\mathcal{\mathcal{P}}_{\rm seed}^{(N)} \big)^N$ and $\mathcal{T}^{(N)}$, $\mathcal{S}^{(N)}$ are defined respectively by~\eqref{eq:defTN} and~\eqref{eq:defSN}.
\end{itemize}
\paragraph{}
These properties rely on the following conjecture (which have been tested for $N=M=2$ and $|\underline{\underline{\lambda}}|\leq 12$):
\begin{tcolorbox}[colback=black!10!white,colframe=black!95!green]
\begin{align}
    Z^{(0)}[\underline{\underline{\lambda}}] =Z[\underline{\underline{\lambda}}] - \sum_{\tiny \begin{matrix} \underline{\alpha}^{(j)}\in \mathcal{S}^{(N)}(\underline{\lambda}^{(j)})\\ j \in \{1,\ldots,N\} \end{matrix}} Z[\underline{\underline{\alpha}}] \cdot Z^{(0)}[\underline{\underline{\lambda}}\ominus \underline{\underline{\alpha}}]= \mathcal{O}(\varepsilon_2^0)\,,\label{StructureAlgM}
\end{align}
\end{tcolorbox}
\noindent
with $\underline{\underline{\lambda}} \ominus \underline{\underline{\alpha}} = \{\lambda^{(i,j)} \ominus \alpha^{(i,j)}\}_{i\in\{1,\ldots,N\},j\in\{1,\ldots,M\}}$.

\paragraph{}
We motivate this conjecture by the following factorisation inherited from~\eqref{eq:prodNp}:
\begin{align}
    &\mathcal{Z}_{\rm def.}^{\rm inst.,(N,M)} = \sum_{\underline{\underline{\lambda}}} \prod_{j=1}^M \mathfrak{q}_j^{|\underline{\lambda}^{(j)}|}  \mathcal{Z}_{\rm bulk}^{\rm inst.,(N,M)}\big|_{\underline{\underline{\lambda}},\varepsilon_2 \to \frac{\varepsilon_2}{N}} \prod_{i=1}^N \prod_{j=1}^M \mathfrak{q}_{i,j}^{k_i^j(\underline{\underline{\lambda}})-|\underline{\lambda}^{(j)}|} \mathcal{Z}_{\rm surf.}^{\rm inst.,(N,M)}\big|_{\underline{\underline{\lambda}}}\,,\nonumber\\
    &\mathcal{Z}_{\rm surf.}^{\rm inst.,(N,M)}\big|_{\underline{\underline{\lambda}}}:= \prod_{l=1}^M \prod_{1\leq i,j\leq N} \prod_{\tiny \begin{matrix}
        p=1\\
        p\neq j-i
    \end{matrix}}^N\frac{\mathcal{N}_{\lambda^{(i,l)}\lambda^{(j,l)}}^{(p|N)}(Q_{a_j^l}/Q_{a_i^{l}},\Qr;q_{1},\tq)}{ \mathcal{N}_{\lambda^{(i,l-1)}\lambda^{(j,l)}}^{(p|N)}(Q_{\widehat{S}} Q_{a_j^l}/Q_{a_i^{l-1}},\Qr;q_{1},\tq)}\,,
\end{align}
we observe that $\mathcal{Z}_{\rm surf.}^{\rm inst.,(N,M)}\big|_{\underline{\underline{\lambda}}}=\mathcal{O}(\varepsilon_2^p)$ with $p\geq 0$. The pole structure in the NS limit is then dominated by the bulk factorised contribution which factorises and lead to a restricted sum over limit shape contributions~\cite{Lee:2020hfu}.

\section{Conclusion and Outlook}\label{sec:conclusion}
\paragraph{}
In this work we study (orbifolds of) $A$-type Little String Theories (LSTs) in the presence of a full type surface defect. The LSTs are constructed as the six-dimensional world-volume theory of $N$ M5-branes on a circle (called $\mathbb{S}_\perp^1$), probing a transverse $\mathbb{R}^4/\mathbb{Z}_M$ (see Table~\ref{tab:m5config} for the details). Below a scale set by the radius of $\mathbb{S}_\perp^1$, this theory resembles a circular quiver gauge theory with $M$ nodes of $U(N)$ and hypermultiplet-matter in the bi-fundamental representation (or adjoint in the case $M=1$), as shown in Figure~\ref{fig:Amquiver}. In this setup, the defect is introduced through a further $\mathbb{Z}_N$ orbifold action, maximally breaking the gauge nodes into $[U(1)]^N$. Using this geometric description, we develop an ADHM construction\footnote{In Appendix~\ref{sec:voa} we provide an alternative (but equivalent) re-derivation in terms of techniques related to vertex operator algebras, reviewed in \cite{Kimura:2023bxy}.} to calculate the full (\emph{i.e.} perturbative and non-perturbative contributions) BPS defect partition function $\mathcal{Z}_{\rm def.}^{\rm inst.}$, which is given in (\ref{eq:defZ}) for $M=1$ and in (\ref{ZinstM}) for generic $M\geq 1$, which to our knowledge was not previously known in the literature. Moreover, both expressions are formulated in a combinatorial fashion, \emph{i.e.} as summations over integer partitions and functions which depend on combinatorial properties of them. This provides a very concrete and compact way of writing the defect partition functions as a function of all physical moduli, which lends itself to further studies.

Indeed, we use the explicit expressions obtained from the ADHM construction to first study non-perturbative symmetries of the defect Little Strings. We find that the symmetries previously found in \cite{Bastian:2018jlf,Filoche:2023yfm} are not broken in the presence of the defect and still leave $\mathcal{Z}_{\rm def.}^{\rm inst.}$ invariant. Furthermore, we analyse the Nekrasov-Shatashvili limit of the defect system: indeed in the na\"ive limit of $\varepsilon_2\to 0$ (which is geometrically defined in the M-brane setup in Table~\ref{tab:m5config}),  $\mathcal{Z}_{\rm def.}^{\rm inst.}$ contains singular contributions. In the literature, two different proposals exist for regularising the latter by normalising $\mathcal{Z}_{\rm def.}^{\rm inst.}$ by suitable (moduli dependent) factors, namely (\ref{eq:regbulk}) and (\ref{eq:curlyZ}), respectively in the case $(M,N)=(1,2)$. We argue that both prescriptions provide a well-behaved NS-limit, due to the recursive relation (\ref{eq:bilin1}) (and its generalisations (\ref{eq:defSN}) to $N>2$ and (\ref{StructureAlgM}) for $M>1$) relating different contributions of the instanton partition functions and which lead to a systematic cancellation of poles. In fact, based on this argument, infinitely many more normalisers are possible by adding additional, precisely specified subsectors of instanton contributions. It will be interesting in the future to explore potential physical interpretations of such modified regularisations. Furthermore, inspired by their structural form, we plan to explore a possible connection between the recursive structures (\ref{eq:bilin1}), (\ref{eq:defSN}), (\ref{StructureAlgM}) and the blow-up equation \cite{Kim:2023glm,Jeong:2020uxz}.

Our work provides explicit expressions for the full BPS defect partition function of a very general class of LSTs. This class encompasses many other theories (in lower dimensions) through particular limits in its parameter space. Indeed, an overview over some of them is systematically given in Table~\ref{tab:rtertesystems}. The explicit form of our defect partition function (notably in its very concrete combinatoric form) therefore gives direct access to these theories as well. Furthermore, (some of) the algebraic structures and (non-perturbative) symmetries are expected to directly percolate to these theories. This is in particular relevant in view of further dualities to various types of integrable models (schematically, see again Table~\ref{tab:rtertesystems}). As a first application of this idea, we have shown in Section~\ref{Sect:BulkHamilton} that for $(M,N)=(1,2)$, the instanton partition function in the bulk decoupling limit takes the form of an elliptic hypergeometric function, which is indeed an eigenfunction of the Hamiltonian of the dual elliptic-Ruijsenaars-Schneider system~\cite{Koroteev:2019gqi}. We expect that in a similar way it is possible to interpret (limits of) our defect partition function (for general values of $(M,N)$) as wave functions of different integrable models. This opens the window to use the non-perturbative symmetries and dualities established on the gauge theory side, for the study of integrable models. In this context it will particularly be interesting to study implications of dualities of the LSTs with $(M,N)$ and $(M',N')$ for $MN=M'N'$ and $\text{gcd}(M,N)=\text{gcd}(M',N')$ as established in \cite{Bastian:2017ary,Bastian:2018dfu}.

In the future it will also be interesting to analyse the defect partition functions from another perspective. 
Indeed, in the M-theoretic setup, the codimension-2 surface defect can be realised by introducing defect M5-branes~\cite{Mori:2016qof} as described in the left part of Table~\ref{tab:braneconstructioncodim2surfacedefect}. This setup is dual to a type IIB fivebrane web in presence of defect D3-branes corresponding to the setup given in the right part of Table~\ref{tab:braneconstructioncodim2surfacedefect}. However, it is not immediately obvious how to directly relate this setup to the realisation using the partial orbifold technique that we have discussed in this paper. It would therefore be interesting to investigate further this connection between the two realisations.
\begin{table}[htbp]
    \centering
    \begin{tabular}{c||c|c|c|c|c|c}
         & $\mathbb{C}_1$ &$\mathbb{C}_2$ & $\mathbb{T}^2$ & $\mathbb{S}^1_{\perp}$ & $\mathbb{C}_3$ & $\mathbb{C}_4$  \\
         \hline\hline
       $N$ M$5$  & -\,- & -\,- & -\,- & $\times$ & &\\
       $k$ M$2$ &  & & -\,- & - & & \\
       $s$ M$5^\prime$ &  & -\,- & -\,- & & & -\,- \\
     \end{tabular}
     \quad 
     \begin{tabular}{c||c|c|c|c|c|c}
         & $\mathbb{C}_1$ &$\mathbb{C}_2$ & $\mathbb{S}^1$ & $\mathbb{S}^1_{\perp}$ & $\mathbb{S}^1_{\rm TN}$ & $\mathbb{R}^3$  \\
         \hline\hline
       $N$ D$5$ &-\,- & -\,-  & - & $\times$ & - &\\
       1 NS$5$ &-\,- & -\,-  & - & - & $\times$ &\\
       $(1,1)$-branes & -\,- & -\,-  & - & - & - &\\
       $k$ F$1$ &  & & -  & - & & \\
       $s$ D$3^\prime$ & & -\,-  & - & & & - \\
     \end{tabular}
    \caption{On the left the brane construction of the codimension-2 surface defect in M-theory as a system of M$5$-M$2$-M$5^\prime$-branes. On the right the dual type IIB setup in which the defect corresponds to the presence of D$3^\prime$-branes.}
    \label{tab:braneconstructioncodim2surfacedefect}
\end{table}

Another potential direction is to further analyze the algebraic structure of the LST system studied in this paper.
As is briefly discussed in Appendix~\ref{sec:voa}, the LST partition function with the defect has an algebraic realization, i.e., the corresponding partition function is realized as a correlation function of the vertex operators. 
Such a correspondence is in general called BPS/CFT correspondence~\cite{Nekrasov:2015wsu}, and the AGT correspondence is one of the well-known examples of this type of correspondence~\cite{Alday:2009aq}, where the codimension-2 defect insertion implies the degenerate field insertion~\cite{Alday:2009fs} or the change of the underlying conformal algebra~\cite{Alday:2010vg}.
Meanwhile, the algebraic construction shown in Appendix~\ref{sec:voa} is rather based on another type of approach, called the quiver W-algebra~\cite{Kimura:2015rgi,Kimura:2016dys,Kimura:2017hez}.
Although this formalism is essentially related to the AGT relation via the type IIB S-duality, a precise relation at the algebraic level is not yet known.
From this point of view, it would be an interesting direction to study the defect LST partition function in terms of the original AGT perspective.
We leave this scope for future work.

\section*{Acknowledgements}
We thank Elli Pomoni, Yegor Zenkevich and Minsung Kim for many enlightening discussions on supersymmetric gauge theories and quantum integrable systems. This project has received financial support from the CNRS through the MITI interdisciplinary programs. The work of TK was in part supported by EIPHI Graduate School (No. ANR-17-EURE-0002) and Bourgogne-Franche-Comté region. BF and SH are grateful to the Institut de Math\'ematiques de Bourgogne as well as the Quantum Theory Center (QTC) at the Danish Institute for Advanced Study and IMADA of the University of Southern Denmark for kind hospitality, while part of this work was performed. BF would furthermore like to thank the DESY Theory group and the group of Piljin Yi at Korean Institute for Advanced Study for hospitality and gratefully acknowledges support from an 'Aide à la mobilité doctorale' of the Université Claude Bernard Lyon 1 and Les Houches Summer School.

\appendix

\section{Definitions and Properties}\label{sec:def}

\subsection{Definitions of modular objects}
This appendix regroups useful definitions related to modular objects used throughout the text. We define the Pochhammer symbol as:
\begin{align}
    (Q_x;Q_\rho)_n := \prod_{i=0}^{n-1}(1-Q_x Q_\rho^i)\,,
\end{align}
using this definition we define the Dedekind $\eta$ function as:
\begin{align}\label{eq:appetadef}
    \eta(\rho) := \Qr^{\frac{1}{24}} (\Qr;\Qr)_\infty\,,
\end{align}
the $\vartheta$ theta function:
\begin{align}\label{eq:appthetedef}
    \vartheta(Q_x;\Qr):= \exp \left(-\sum_{n\in\mathbb{Z}^*} \frac{1}{n} \frac{Q_x^{n}}{1-\Qr^n}\right)=(Q_x^{-1} \Qr;\Qr)_\infty (Q_x;\Qr)_\infty\,,
\end{align}
which satisfies the quasi-periodicity condition:
\begin{align}\label{eq:appquasitheta}
    \vartheta(Q_x\Qr^{-1};\Qr)= \Qr^{-1} Q_x \,\vartheta(Q_x;\Qr),
\end{align}
and is related to the $\theta_1$ Jacobi theta function by:
\begin{align}\label{eq:appcurlythetajacobi}
    \theta_1(x;\rho)=i \Qr^{\frac{1}{12}} Q_x^{-\frac{1}{2}} \,\eta(\rho)\, \vartheta(Q_x;\Qr)\,.
\end{align}
We then define the elliptic Pochhammer symbol and its multivariable generalisations as:
\begin{align}\label{eq:appellpochdef}
    &\vartheta(Q_x;q)_n:= \prod_{i=0}^{n-1} \vartheta(Q_xq^i;\Qr)\,, && \vartheta(a_1,\ldots,a_k;q)_n := \prod_{i=0}^{n-1} \vartheta(a_i;q)_n\,,\nonumber\\
    &\vartheta(a;q_1,q_2)_n:= \prod_{i=0}^{n-1} \prod_{j=0}^{n-1} \vartheta(a q_1^i q_2^j;\Qr)\,.
\end{align}
Using this definition, we define the $N$ arguments elliptic hypergeometric function $_NE_{N-1}$ as:
\begin{align}\label{eq:appellhyp}
    \,_NE_{N-1} \left[\begin{matrix}
        a_1 & a_2 & \cdots & a_N \\ \cdot & b_1 & \cdots & b_{N-1}
    \end{matrix}; q, \Qr ; \mathfrak{q} \right] =\sum_{n=0}^\infty \mathfrak{q}^n \frac{\vartheta(a_1,\ldots,a_N;q)_n}{\vartheta(q;q)_n \vartheta(b_1,\ldots,b_{N-1};q)_n} \,,
\end{align}
with the balance condition $a_1 a_2 \cdots a_N = q b_1 b_2 \cdots b_{N-1}$.

\subsection{Some properties of $\omega_{k|N}$}~\label{subsec:propome}

As detailed in subsection~\ref{subsec:combexprM=1}, the combinatoric expression of the defect instanton PF can be easily obtained using a projection on characters $\omega_{k|N}$ defined by~\eqref{eq:defomega}. We now give some of the key properties $\omega_{k|N}$ satisfies which are useful for our computation. Naturally, $\omega_{k|N}$ is a projection on the space of Laurent series $\mathbb{C}((\tq))$ and satisfies:
\begin{align}\label{eq:genpropome}
    \omega_{k_1|N} \circ \omega_{k_2|N} = \delta_{k_1,k_2}\omega_{k_2|N}\,, && \sum_{k=1}^N \omega_{k|N}=\mathds{1} \,, && \omega_{k+N|N} = \omega_{k|N}\,,
\end{align}
a $\tq$ factor can by pulled out of $\omega_{k|N}$ when compensated by a shift of $k$:
\begin{align}\label{eq:shiftpropome}
    \forall P \in \mathbb{C}((\tq))\,, && \omega_{k|N}(\tq \cdot P) = \tq \cdot \omega_{k-1|N}(P)\,,
\end{align}
which leads to the following product property $\forall n,m \in \mathbb{Z}$:
\begin{align}\label{eq:prodome}
    \forall P,Q \in \mathbb{C}((\tq))\,, \quad\omega_{n|N}(P)\cdot \omega_{m|N}(Q) = \omega_{n+m|N}(\omega_{n|N}(P)\cdot Q)=\omega_{n+m|N}(P\cdot \omega_{m|N}(Q))\,.
\end{align}
Under the dual character operation defined by~\eqref{eq:dualoperation}, we have:
\begin{align}\label{eq:dualpropome}
    \forall P \in \mathbb{C}((\tq))\,,\quad\quad \forall k \in \mathbb{Z}\,,\quad\quad\quad \left( \omega_{k|N}(P)\right)^\vee=\omega_{-k|N}(P^\vee)\,.
\end{align}

\section{Vertex operator algebraic derivation}\label{sec:voa}
\paragraph{}
In this Appendix, we discuss the vertex operator algebra derivation of the defect PF. This construction allow to easily connect the brane construction with the algebraic computation of \cite{shiraishi2019affinescreeningoperatorsaffine}. In particular, we will detail the realisation in term of a free field algebra and then reformulate the defect PF using screening currents. For more detail on these constructions we refer the reader to \cite{Kimura:2022zsm,Kimura:2023bxy}.
\subsection{Free field realisation}
We first discuss the free field realisation of the surface defect. We perform a dimensional reduction to a type IIB setup, the M-theoretic brane system reduces to a D$1$-D$5$ system. We define the Heisenberg algebra for the D$1$-strings on the $\mathbb{C}_1 \times \mathbb{C}_2/\mathbb{Z}_N \times \mathbb{C}_3 \times \mathbb{C}_4/\mathbb{Z}_N \times \mathbb{T}^2$, with the orbifold action defined by~\eqref{eq:orbifoldactionM1}, as:
\begin{align}
    [{\sf a}^{i,(\pm)}_{n},{\sf a}^{j,(\pm)}_{m}] = \mp \frac{1}{n} [(1- q_1^{\pm n})(\mathds{1}_N-\tq^{\pm n} \mathcal{R}_1])(1-q_3^{\pm n})(\mathds{1}_N - q_4^{\pm n}\mathcal{R}_{N-1})]_{ij} \frac{\delta_{n+m,0}}{1-\Qr^{\pm n}}\,,
\end{align}
where the indices $i$ and $j$ indicate sub-sectors under the $\mathbb{Z}_N$ orbifold and $[\cdot]_{ij}$ indicated the $(i,j)$ matrix entry. We can then define the algebra associated with D$3$-branes spanning the $\mathbb{C}_1$ plane:
\begin{align}\label{eq:defs1}
    &{\sf s}_{1,n}^{i,(\pm)} = \frac{{\sf a}^{i,(\pm)}_{n}}{1- q_1^{\mp n}}\,,\nonumber\\
    & [{\sf s}_{1,n}^{i,(\pm)},{\sf s}_{1,m}^{j,(\pm)}] = \mp\frac{1}{n} \frac{[(\mathds{1}_N-\tq^{\pm n} \mathcal{R}_1)(1-q_3^{\pm n})(\mathds{1}_N - q_4^{\pm n} \mathcal{R}_{N-1})]_{ij}}{1-q_1^{\mp n}} \frac{\delta_{n+m,0}}{1-\Qr^{\pm n}}\,,
\end{align}
similarly the D$5$-brane algebra spanning $\mathbb{C}_1 \times \mathbb{C}_2$ is given by:
\begin{align}
    &{\sf x}_{12,n}^{i,(\pm)} = \sum_{j=1}^N \left[(\mathds{1}_N - \tq^{\mp n} \mathcal{R}_1)^{-1}\right]_{ij} \cdot {\sf s}_{1,n}^{j,(\pm)}\,,\nonumber\\
    & [{\sf x}_{12,n}^{i,(\pm)},{\sf x}_{12,m}^{j,(\pm)}] = \mp \frac{1}{n} \left[\frac{(1-q_3^{\pm n})(\mathds{1}_N - q_4^{\pm n} \mathcal{R}_{N-1})}{(1- q_1^{\mp n})(\mathds{1}_N - \tq^{\mp n} \mathcal{R}_1)}\right]_{ij} \frac{\delta_{n+m,0}}{1-\Qr^{\pm n}}\,,\nonumber\\
    &[{\sf a}_n^{i,(\pm)},{\sf x}_{12,m}^{j,(\pm)}]= \mp \frac{1}{n} \left[ (1-q_3^{\pm n})(\mathds{1}_N - q_4^{\pm n} \mathcal{R}_{N-1})\right]_{ij}\frac{\delta_{n+m,0}}{1-\Qr^{\pm n}}\,,\nonumber\\
    &[{\sf x}_{12,m}^{i,(\pm)},{\sf a}_n^{j,(\pm)}]=\pm \frac{1}{n} \left[\frac{q_1^{\pm n}(\mathds{1}_N - \tq^{\pm n}\mathcal{R}_1)(1-q_3^{\pm n})(\mathds{1}_N - q_4^{\pm n}\mathcal{R}_{N-1})}{(\mathds{1}_N - \tq^{\mp n}\mathcal{R}_1)}\right]_{ij} \frac{\delta_{n+m,0}}{1-\Qr^{\pm n}}\,.
\end{align}
We then define their associated vertex operators:
\begin{align}
    &{\sf A}_i(x) = {\sf a}_0^i(x) : \exp \left(\sum_{n \in \mathbb{Z}^*} ({\sf a}_n^{i,(+)} x^{-n}+{\sf a}_n^{i,(-)} x^n) \right)\!:\,,\nonumber\\
    &{\sf X}_i(x) = {\sf x}_0^i(x) : \exp \left(\sum_{n \in \mathbb{Z}^*} ({\sf x}_n^{i,(+)} x^{-n}+{\sf x}_n^{i,(-)} x^n) \right)\!:\,,\nonumber
\end{align}
where ${\sf a}_0^i(x),{\sf x}_0^i(x)$ are zero-mode contributions which are related to the fractional gauge coupling (more details about the zero-mode contributions can be found in \cite{Kimura:2023bxy}). The LMNS integral formula~\eqref{eq:LMNS} can then be obtained by considering the following contribution:
\begin{align}
    \mathcal{Z}_{\rm def.}^{\underline{k},N}= \frac{1}{k!} \prod_{i=1}^N \prod_{I=1}^{k_i} {\rm d}\phi_{i,I} \left\langle0 \left|\prod_{i=1}^N \prod_{I=1}^{k_i} {\sf A}_i^{-1}(\phi_{i,I}):\prod_{j=1}^N \prod_{\alpha=1}^N {\sf X}_j(a_{i,\alpha}):\right|0 \right\rangle\,,
\end{align}
where $|0\rangle$ is defined as the vacuum of the Fock space generated by ${\sf a}^{i,(\pm)}_n$, and the perturbative contribution is given by ${\sf XX}$ correlators:
\begin{align}
    \mathcal{Z}_{\rm def.}^{\rm pert.,(N)} =  \left\langle0 \left|\prod_{j=1}^N \prod_{\alpha=1}^N {\sf X}_j(a_{i,\alpha})\right|0 \right\rangle\,.
\end{align}

\subsection{Screening charges}
After integration, we can deduce from the result of subsection~\ref{subsec:combexprM=1} an alternative presentation contribution of the fixed point characterised by $\underline{\lambda}$ to the defect instanton PF. For all $i \in \{1,\ldots,N\}$, we define:
\begin{align}
    \mathcal{X}_{\lambda^{(i)}}:= \left\{Q_{a_i} q_1^{\lambda^{(i)}_l} \tq^{l-1} \right\}_{l \in \mathbb{N}^*}\,, && \omega_{k|N} (\mathcal{X}_{\lambda^{(i)}}):= \left\{Q_{a_i} q_1^{\lambda^{(i)}_l} \tq^{l-1} \right\}_{l \in \mathbb{N}^*, l \equiv k \!\mod N}\,,
\end{align}
where the set of fixed point is characterised by $\{ \omega_{k|N}(\mathcal{X}_{\lambda^{(i)}}) \}_{i,k \in \{1,\ldots,N\}}$. We then defined the screening charges as:
\begin{align}
    &{\sf S}_{i,j}[\underline{\lambda}] := \prod_{x \in \omega_{j|N}\big(\mathcal{X}_{\lambda^{(i-j+1)}}\big)}^{\prec}:\exp \left( \sum_{m\in \mathbb{Z}^*} \mathsf{s}_{1,m}^{k,(+)} x^{-m} + \mathsf{s}_{1,m}^{k,(-)} x^{m}  \right):\,, \\
    &{\sf S}_{i}[\underline{\lambda}] := \left[\prod_{1\leq j \leq i}^\prec {\sf S}_{i,j}\right] \left[\prod_{i+1\leq j \leq N}^\prec {\sf S}_{i,N+j} \right]\,,
\end{align}
we use the symbol $\prod_{i=1,\ldots,n}^\prec a_i=a_1 a_2 \cdots a_n$ to indicate the operator ordering, ${\sf s}_{1,m}^{i,(\pm)}$ is the generator associated with a $D3$-brane~\eqref{eq:defs1}. The full partition function can then be obtained as:
\begin{align}
    \mathcal{Z}_{\rm def.} = \sum_{\underline{\lambda}} \left\langle0 \left| \prod_{i=1}^N \mathfrak{q}_i^{k_i(\underline{\lambda})}\, {\sf S}_i[\underline{\lambda}] \right|0 \right\rangle\,.
\end{align}




\bibliographystyle{utphys}
\bibliography{biblio2}

\end{document}